\documentclass[apj]{emulateapj}
\usepackage{graphicx}

\usepackage{color}
\usepackage{graphicx}
\usepackage{amsmath,amssymb}
\usepackage[utf8]{inputenc}
\usepackage[T1]{fontenc}

\def\beq{\begin{equation}}
\def\enq{\end{equation}}
\def\bea{\begin{eqnarray}}
\def\ena{\end{eqnarray}}

\def\Ek525{E_{k,52.5}}

\begin{document}


\title{High-Energy Neutrino Flares from X-Ray Bright and Dark Tidal Disruption Events}

\author{Nicholas Senno\altaffilmark{1,2,3}, Kohta Murase\altaffilmark{1,2,3,4}, and Peter M\'esz\'aros\altaffilmark{1,2,3}}
\altaffiltext{1}{Department of Physics, The Pennsylvania State University, University Park, PA 16802, USA}
\altaffiltext{2}{Department of Astronomy \& Astrophysics, The Pennsylvania State University, University Park, PA 16802, USA}
\altaffiltext{3}{Center for Particle and Gravitational Astrophysics, The Pennsylvania State University, University Park, PA 16802, USA}
\altaffiltext{4}{Yukawa Institute for Theoretical Physics, Kyoto University, Kyoto 606-8502, Japan}

\date{\today}

\begin{abstract}
X-ray and $\gamma$-ray observations by the {\it Swift} satellite revealed that a fraction of tidal disruption events (TDEs) have relativistic jets.  Jetted TDEs have been considered as potential sources of very high-energy cosmic-rays and neutrinos.  In this work, using semi-analytical methods, we calculate neutrino spectra of X-ray bright TDEs with powerful jets and dark TDEs with possible choked jets, respectively.  We estimate their neutrino fluxes and find that non-detection would give us an upper limit on the baryon loading of the jet luminosity contained in cosmic-rays  $\xi_{cr} \lesssim20-50$ for Sw J1644+57. We show that X-ray bright TDEs make a sub-dominant ($\lesssim5-10$\%) contribution to IceCube's diffuse neutrino flux, and study possible contributions of X-ray dark TDEs given that particles are accelerated in choked jets or disk-winds.  We discuss future prospects for multi-messenger searches of the brightest TDEs.  
\end{abstract}

\pacs{none \vspace{-0.3cm}}

\maketitle


\section{Introduction}
Supermassive black holes (SMBHs) with masses $\gtrsim 10^5 M_\odot$ can disrupt stars whose orbits pass within the tidal disruption radius \citep[see a review][and references therein]{2015JHEAp...7..148K}.  
For a SMBH with $M_{\rm BH,6}  = 10^6\,M_\odot$ disrupting a Sun-like star, this distance is $r_T\approx 10^{13}\,{\rm cm}\,M_{\rm BH,6}^{1/3}\,M_\odot^{-1/3}\,R_\odot$. Roughly half the disrupted stellar material becomes unbound from the SMBH, while the remaining plasma circularizes to form an accretion disk~\citep{Rees:1988ei,1989IAUS..136..543P,1989ApJ...346L..13E}. It has been argued that material from a tidal disruption event (TDE) would accrete at a super-Eddington rate $\dot{M}_{\rm Edd} \sim 2\times10^{-3}\,M_\odot/{\rm yr}$~\citep[e.g.,][]{2009MNRAS.400.2070S,2013ApJ...767...25G,2013MNRAS.435.1809S,2015ApJ...812L..39D}, and potentially launch a jet (e.g., through the Blandford-Znajek mechanism) \citep[e.g.,][]{2012ApJ...760..103D,2012ApJ...749...92K,2014MNRAS.437.2744T,Piran:2015gp,2015MNRAS.454L...6M}. 

Relativistic jets have been promising candidates of high-energy particle accelerators, as commonly considered in the literature of gamma-ray bursts (GRBs) and active galactic nuclei (AGN).  The X-ray detections of jetted TDEs  \citep{Burrows:2011kz,2012ApJ...753...77C,Brown:2015du} have stimulated proposals of TDEs as potential ultrahigh-energy cosmic-ray (UHECR) accelerators \citep{2009ApJ...693..329F,2014arXiv1411.0704F} and high-energy cosmic neutrino sources~\footnote{\cite{2008AIPC.1065..201M,MT09} calculated the diffuse neutrino intensity from TDEs, based on the giant flare scenario suggested by \cite{2009ApJ...693..329F}.} \citep{2008AIPC.1065..201M,MT09,2011PhRvD..84h1301W}. 
In particular, the recent discovery of astrophysical neutrinos by IceCube (IC) \citep{ICdiscovery,ICevidence} gives us a good motivation to revisit various astrophysical transients, including TDEs, as multi-messenger sources.  The sources of the very high-energy neutrinos (VHE) with energies $E_\nu \gtrsim 10$ TeV are unknown. The arrival directions and flavor composition of these neutrinos are consistent with an isotropic diffuse flux of astrophysical neutrinos of extragalactic origin \citep{2015ApJ...809...98A}.  
Sources such as GRBs \citep{1997PhRvL..78.2292W} and Blazars \citep{1995APh.....3..295M} do not account for a significant fraction of the observed diffuse neutrino flux \citep{IceCubeCollaboration:2016tj,2016arXiv161103874I,2016arXiv161106338N}. 
Cosmic-ray reservoirs such as Starburst Galaxies \citep{2006JCAP...05..003L,2006astro.ph..8699T} and Galaxy Clusters/Groups \citep{2008ApJ...689L.105M,2009ApJ...707..370K} are both neutrino and $\gamma$-ray bright, and can significantly contribute to the diffuse flux \citep[e.g.,][]{2013PhRvD..88l1301M,2014JCAP...09..043T,2015PhLB..745...35C,2015ApJ...806...24S,2015arXiv150900983B,2015ApJ...805...95C,2015A&A...578A..32Z,2016ApJ...828...37F}, but only above 0.1 PeV energies because of stringent $\gamma$-ray constraints \citep{Murase:2016ez}. 

Models of $\gamma$-ray ``hidden'' sources--which mask their $\gamma$-ray emission at $\gtrsim1$ GeV energies--are not constrained by {\it Fermi}-LAT or IC analyses \citep{Murase:2016ck}. Such sources include choked GRB jets~\citep{2013PhRvL.111l1102M,2016PhRvD..93h3003S,2016PhRvD..93e3010T}, newborn pulsars~\citep{2016JCAP...04..010F}, white dwarf mergers~\citep{2016ApJ...832...20X}, and high-redshift galaxies~\footnote{Note that the $\gamma$-ray cutoff energy due to the extragalactic background cannot be lower than $\sim10$~GeV \citep{2012Sci...338.1190A,2013IJMPD..2230025C,2015ApJ...805...33K}.} \citep{2016ApJ...826..133X}.  X-ray bright, jetted TDEs are an example of hidden sources, since high-energy $\gamma$-rays are significantly absorbed via two-photon annihilation \citep{Burrows:2011kz}. Alternatively, if an unbound material or optically-thick wind forms a spherical circumnuclear envelope, it could obscure or reprocess the non-thermal emission from a relativistic jet (i.e., a choked-jet TDE), making such events potential ``hidden'' neutrino sources \citep{2016PhRvD..93h3005W}. 

In this work, we revisit high-energy neutrino production in TDE jets. We calculate neutrino fluxes from both X-ray bright successful TDE jets and possible choked TDE jets that could occur if the jets are not powerful enough. For Sw J1644+57 (hereafter Sw 1644), we also use IC data for upgoing muon neutrino events to search for coincidences between neutrino detection and the three jetted TDE candidates seen by Swift-BAT in 2011\citep{Aartsen:2015de}. While the right ascension and arrival time is not given for these neutrinos, we place meaningful limits on the baryonic loading factor of jetted TDEs. Then, we evaluate diffuse neutrino intensities of jetted TDEs and discuss the present constraints.

\section{X-Ray Bright TDEs with Successful Jets}\label{XTDE}

We first consider jetted TDEs that have non-thermal X-ray spectra ($0.3\lesssim E_\gamma \lesssim 150$ keV) to determine the seed photon density for $p\gamma$ interactions. 
Only Sw 1644 was observed early enough by Swift XRT to fit an SED \citep{Burrows:2011kz}, although two jetted TDE candidates Sw J1112-82 \citep{Brown:2015du} and Sw J2058+05 \citep{2012ApJ...753...77C} show similar peak X-ray luminosities.  We use a log-parabolic fit to the SED of Sw 1644 \citep{Burrows:2011kz}, which is given by
\begin{equation}
\label{eq:energy_dense}
\varepsilon^2 n_\varepsilon =A {(\varepsilon/\varepsilon_{\rm pk})}^{0.5-0.25\hat{a}\log(\varepsilon/\varepsilon_{\rm pk})},
\end{equation}
where $E_{\rm pk}=200$~keV , $\hat{a}$, and $A$ are fitting parameters (see Fig.~1). The peak energy corresponds to $\varepsilon_{\rm pk}=20$~keV in the jet plasma comoving frame for Lorentz factor $\Gamma=10$. If the non-thermal X-rays are produced by synchrotron emission from leptons in the jet, the template from Sw 1644 can be adapted to fit jetted TDEs of different luminosities. 
The luminosity changes with time and the maximum luminosity reaches $L_{\rm max}\equiv\varepsilon L_{\varepsilon}^{\rm pk}\sim{10}^{48}~{\rm erg}~{\rm s}^{-1}$, which lasts for 3 days (implying a duration of $t_{\rm dur}\sim2\times{10}^5$~s in the cosmic rest frame). However, it decreases after the peak, and the median luminosity in the 0.3-10~keV band is $L_{[0.3,10~\rm keV]}=8.5\times{10}^{46}~{\rm erg}~{\rm s}^{-1}$, considering the emission with duration of $t_{\rm dur}\sim10^6$~s.  The corresponding bolometric luminosity is $L_{\gamma}=5.7\times{10}^{47}~{\rm erg}~{\rm s}^{-1}$, and we use the bolometric radiation energy of $L_{\gamma}t_{\rm dur}=5.7\times10^{53}$~erg at $L_\gamma=10^{47}~{\rm erg}~{\rm s}^{-1}$, which is relatively conservative for Sw 1644.  
As we see below, the meson production efficiency ($f_{p\gamma}$) is proportional to $L_\gamma$ so neutrino production is expected to be dominated by the high state that lasts during $t_{\rm high}$. Thus, for estimates of neutrino fluences, we will use $L_{[0.3,10~\rm keV]}^{\rm pk}={10}^{48}~{\rm erg}~{\rm s}^{-1}$ corresponding to $L_{\gamma,\rm pk}\simeq12L_\gamma$.  
In a relativistic jet the turbulent magnetic field energy is parametrized as $B^2/(8\pi)=\xi_B L_{\gamma,\rm pk}/(4\pi r^2\Gamma^2c)$ during the high state, and for TDEs with different luminosities, the peak synchrotron flux goes as $\varepsilon F_{\varepsilon}^{\rm pk} \propto L_{\gamma,\rm pk}$ and $\varepsilon_{\rm pk} \propto B \propto L_{\gamma,\rm pk}^{1/2} r_{\rm em}^{-1}$, where $r_{\rm em}$ is the internal dissipation radius.

The locations of the non-thermal emission from both Sw J1644 and 2058 are believed to be close to the jet base since both show variability with $\delta t \sim 10^2\,{\rm s}$. Assuming -- was inferred for Sw 1644 \cite{Burrows:2011kz} -- that TDE jets are modestly relativistic ($\Gamma\sim10$), the emission radius is estimated to be $r_{\rm em}\sim 3\times10^{14}\,{\rm cm}~\Gamma_1^2{\delta t}_2$ which corresponds to a few hundred Schwarzschild radii from the SMBH \citep{Bloom:2011er} \footnote{See however \cite{2016Natur.535..388K} who argued that the soft X-ray emission may come from the accretion disk and produced $\sim10$ Schwarzschild radii from the SMBH.}. We assume $\Gamma$ and $\delta t$ -- and therefore the internal dissipation radius $r_{\rm em}$ -- are similar for jetted TDEs of all luminosities. 
This location of the emission region is consistent with the observation of fast X-ray variability while optical emission may be produced by the jet's forward shock at a distance of $\sim 10^{15}\,{\rm cm}$ from the SMBH \citep[e.g.,][]{Pasham:2015bj}. As we will show below, X-ray bright jetted TDEs are likely to have successful jets with isotropic equivalent luminosity $L \gtrsim10^{44.5}\,{\rm erg/s}$. 

\begin{figure}
\includegraphics[width=0.45\textwidth]{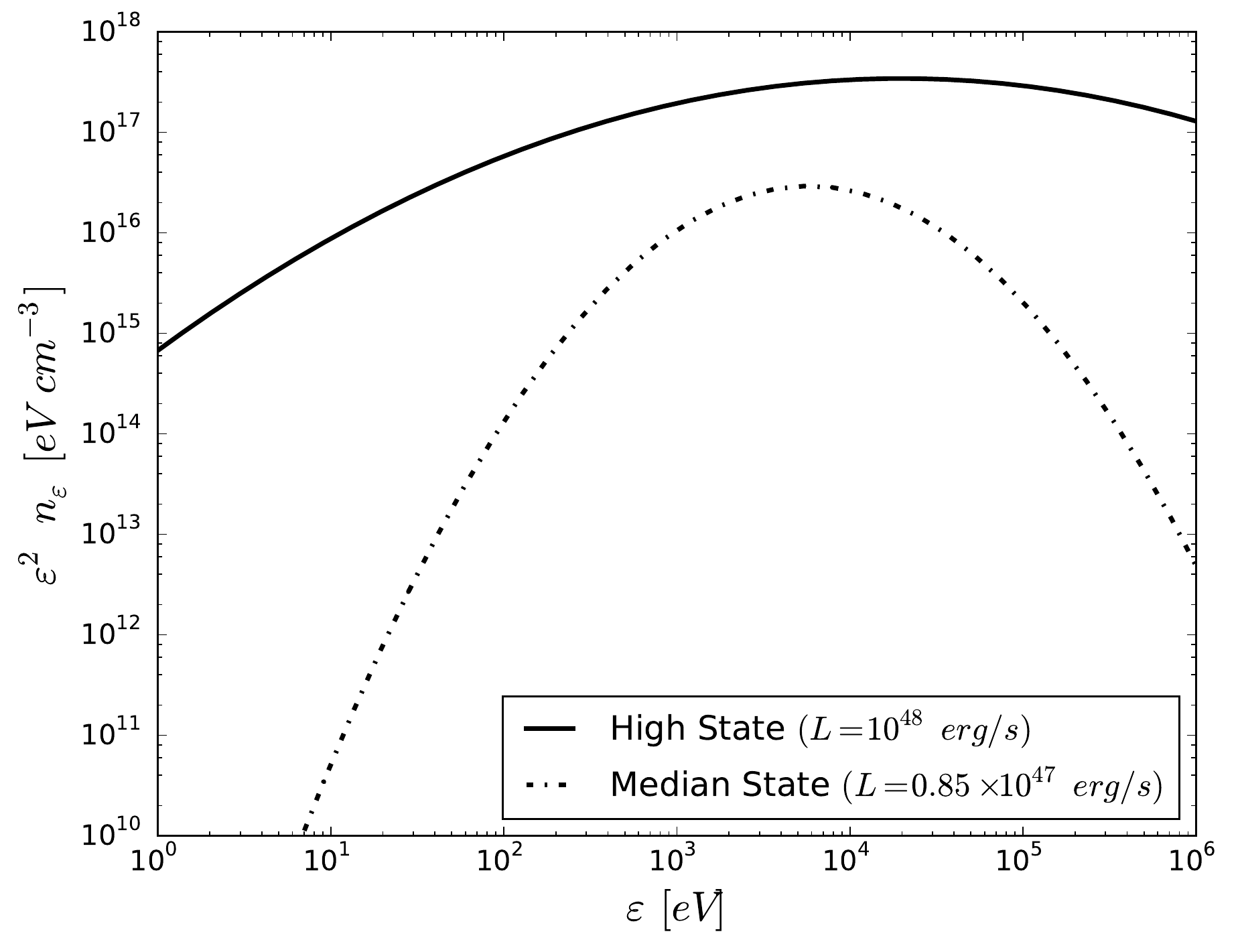}
\caption{The non-thermal SED of Sw 1644 as a function of the comoving energy $\varepsilon$ in the high (solid) and median (dash-dot) luminosity states. One see that there is a peak around 10~keV, corresponding to $E_{\rm pk}=200$~keV.}
\vspace{-1.\baselineskip}
\label{fig:cooling2} 
\end{figure}

\section{X-Ray Dark TDEs with Choked Jets?}
\subsection{Jet Propagation}

It has been generally assumed that TDE jets are powered by the Blandford-Znajek process. However, not all TDEs may have visible jets. The observed rate of X-ray bright jetted TDEs is $\rho_{0,\rm X-TDE}\sim0.03~{\rm Gpc}^{-3}~{\rm yr}^{-1}$, which is much smaller than the total TDE rate at $z=0$, $R_{0,\rm all-TDE}\sim10^3-10^5~{\rm Gpc}^{-3}~{\rm yr}^{-1}$ \citep[see][and references therein]{2016MNRAS.461..371K}. 
This may simply come from the fact that only a fraction of the TDEs may have jets. But, an alternative possibility is that jetted TDEs are more common but many of them are stuck in the TDE material \citep{2016PhRvD..93h3005W}.  It has been speculated that SMBHs could have a thick circumnuclear material of size $r_{\rm out}\sim 3\times10^{15}\,{\rm cm}$ produced by the breakup of the star \citep{1997ApJ...489..573L,2012ApJ...760..103D}. Since the visible TDE emission radius is inside of the circumnuclear envelope, the non-thermal emission will only be visible if the jet breaks out. 
In this scenario, the SMBH is surrounded by an optically thick material formed from the disrupted stellar material that is not bound in an accretion disk ($\sim 0.5M_\ast$). This envelope material may have a wind density distribution with $\varrho_{\rm env}\propto r^{-2}$, in which the material is dominated by disk-driven winds. We consider a density profile proposed by \cite{1997ApJ...489..573L} and \cite{2012ApJ...760..103D},
\begin{equation}
\label{eq:density}
\varrho_{\rm env}(R)  = \frac{f_{\rm TDE}M_\ast}{4\pi\ln(r_{\rm out}/r_{\rm in})r^3}\equiv \varrho_{\rm in}r_{\rm in}^3 r^{-3}
\end{equation}
where the envelope may be defined by inner and outer radii $r_{\rm in}\sim r_T \sim 3\times10^{13}\,{\rm cm}$ and $r_{\rm out}\sim3\times10^{15}\,{\rm cm}$  and $\varrho_{\rm in}=6.4\times{10}^{-10}~{\rm g}~{\rm cm}^{-3}$. 
Correspondingly, the radiation temperature is assumed to be
\begin{equation}
T_{\rm env}=10^{6}~{\rm K}~{\left(\frac{r}{r_{\rm in}}\right)}^{-1}.
\end{equation}

The envelope material changes the dynamics of the jet launched by the SMBH, in a way that is similar to the star/jet interaction in a long GRB in the collapsar scenario \citep{2011ApJ...740..100B,2013ApJ...777..162M}. Specifically, the dimensionless velocity of the jet head can be related to the isotropic equivalent jet kinetic luminosity $L$ and the density of the envelope \citep{2011ApJ...740..100B} where
$$
\tilde{L}\approx \frac{L}{4\pi r^2\varrho_{\rm env}c^3}.
$$
When $\tilde{L} \ll \theta_j^{-4/3}$ -- as it is for the case of choked jet TDEs -- collimation shocks are formed and they change the jet's initially conical shape to a cylindrical one. 
The position of the jet head approximates the forward position of the collimated jet, which is given by \citep{2013ApJ...777..162M}
\begin{equation}
r_h\approx2.5\times{10}^{15}~{\rm cm}~t_{\rm eng,6}^{3/2}L_{44}^{1/2}\varrho_{\rm in}^{-1/2}r_{\rm in,13.5}^{-3/2}\theta_{j,-1}^{-1},
\end{equation}
assuming that the jet is powered by the accretion disk for a time $t_{\rm eng} \sim 10^6\,M_{\rm BH,6}^{1/2}\,{\rm s}$, corresponding to the period of the most tightly bound material in the accretion disk \citep{2012ApJ...749...92K}. 
Using the assumed value for $t_{\rm eng}$ and the relationship above for the position of the jet head, we find that only weak jets with luminosity 
\begin{equation}
L\lesssim 2\times10^{44}\,{\rm erg/s}~t_{\rm eng,6}^{-3}\varrho_{\rm in}r_{\rm in,13.5}^{3}r_{\rm out,15.5}^2\theta_{j,-1}^{2},
\end{equation}
can be choked by the envelope (i.e., $r_h(t_{\rm eng}) \lesssim r_{\rm out}$).  
Assuming the fraction of jet luminosity converted to photons $\epsilon_\gamma\simeq0.2$ is the same for all jets, we extrapolate the luminosity function of \cite{2015ApJ...812...33S} to estimate the number of jetted ``hidden'' jetted TDEs which occur with jet luminosities $L\lesssim 10^{44.5}\,{\rm erg/s}$. Note that the existence of jets with $L_j\approx(\theta_j^2/2)L\lesssim{10}^{42}~\theta_{j,-1}^2~{\rm erg}~{\rm s}^{-1}$ (that is a sub-Eddington luminosity) is not guaranteed, and numerical simulations have not found such weak jets in the current setup~\citep{2014MNRAS.437.2744T,2015MNRAS.454L...6M}. 
If the cocoon pressure is assumed to be constant, the collimation shock radius can be as large as $r_{\rm cs}\sim r_h$. However, as indicated in \cite{2013ApJ...777..162M}, a pressure gradient exists in more realistic situations (especially if $r_{\rm cs}\ll r_{h}$), and the collimation shocks may occur near the inner envelope radius $r_{\rm cs}\sim r_{\rm in}$. In this work, in order to discuss an optimistic case, we assume that the collimation shock radius is large enough that internal dissipation occurs at $r_{\rm em}\sim3\times{10}^{14}~{\rm cm}$.


\subsection{Photon Field Modeling}
\label{ss:photons}

For bright jetted TDEs such as Sw 1644, the jet's non-thermal spectrum, and therefore the target photons for $p\gamma$ interactions are readily available from observations. Choked jets by definition have their photon fields hidden from us. 
Thus, we must extrapolate their SEDs under several assumptions. 

In this work, we consider two emission regions. The first is the internal shock site, which can occur near the collimation shock. This scenario is an extrapolation of the X-ray bright TDE jet scenario. In addition to non-thermal photons produced inside the jet, we consider thermal radiation fields provided by the external material. 
The second is the termination shock, which exists around the jet head. This is unique to choked jet models.  However, if the shock is radiation-mediated, the diffusive shock acceleration of cosmic-rays is inefficient so that high-energy neutrino production is not expected \citep{2013PhRvL.111l1102M}. In the TDE cases, both internal and termination shocks are collisionless and radiation-unmediated, so that one may expect efficient particle acceleration, in principle, if choked jets exist. Note that the meson production efficiency in such hidden cosmic-ray accelerators is always high so that the system is regarded as calorimetric. Cosmic-ray acceleration does not have to be located inside jets. One may consider disk-driven winds as alternative cosmic-ray acceleration sites. Also, observations suggested that outflows are likely to have a complicated structure \citep{2015MNRAS.450.2824M}. 

In the internal shock scenario, the shock radius is assumed to be $r_{\rm em}=3\times{10}^{14}$~cm. 
We have two radiation fields. One is thermal radiation from the envelope, which has temperature of $T=10^5$~K at this radius. We checked that thermal radiation from the hot cocoon is sub-dominant in our setup.  However, for the radiation from the optically-thick envelope, only a fraction of $f_{\rm esc}\approx1/\tau_T\simeq1.2\times{10}^{-2}~\varrho_{\rm in}^{-1}r_{\rm in,13.5}^{-3}r_{14.5}^2$ can escape to the optically-thin jet region \citep{2013PhRvL.111l1102M}. Taking into account that the thermal photon number density is boosted by $\Gamma$ in the jet comoving frame, the comoving thermal photon density in the jet is
\begin{equation}
n_{\gamma, \rm is}^{\rm th}\approx\Gamma f_{\rm esc}16\pi \zeta(3){(kT/hc)}^3.
\end{equation}
For the non-thermal synchrotron spectrum, we again use a template synchrotron spectrum of Sw 1644 to estimate the target photon density in lower-luminosity jetted TDEs \citep{Burrows:2011kz}, assuming $E_\gamma^{\rm pk}\propto L_\gamma^{1/2} r_{\rm em}^{-1}$. Note that the synchrotron photons are produced by electrons co-accelerated in the internal shocks and are not significantly modified. See \S \ref{XTDE}.

The second scenario is the termination shock scenario, in which cosmic-rays are accelerated at the termination shock caused by the cylindrical jet with $\Gamma_{\rm cj}\approx\theta_j^{-1}$. Note that, in the GRB case, this scenario does not work because the termination shock is radiation-mediated and efficient particle acceleration would not occur. However, in the TDE case, the shock is collisionless since the jet density is small enough.  The blackbody spectrum from the jet head is estimated using the photon energy density $U_{\gamma h}\approx L/(4\pi\Gamma_h^2r_{h}^2c) \simeq 84\,L_{44.5}\Gamma_h^{-2}r_{h,15.5}^{-2}\,{\rm erg\,cm^{-3}}$ and the relationship $U_{\gamma h}=aT_h^4$, where the jet head temperature is $T_h \simeq 1.0\times10^4~L_{\gamma,44}^{1/4}\Gamma_h^{-1/2}r_{h,15.5}^{-1/2}\,{\rm K}$ \citep{2001PhRvL..87q1102M,2003PhRvD..68h3001R}. 
The jet head is non-relativistic and $\Gamma_h\sim 1$ ($\beta_h\ll1$) is obtained.  Assuming the thickness of the jet head is $\Delta r_{h} \approx r_{h}/\Gamma_{\rm cj}\sim0.1 r_h$, the jet is found to be optically-thin with $\tau_h \approx n_h \sigma_T \Delta r_h <1$, so we do not have to include the suppression factor in this estimate. The photon density is simply estimated to be $n_{\gamma,\rm ts}^{\rm th}\approx16\pi \zeta(3){(kT_h/hc)}^3$.  Note that the circumnuclear envelope is still optically thick and manages to absorb and scatter any $\gamma$-rays so the system is hidden in $\gamma$-rays.

\section{Detection of neutrinos from individual TDEs}
\subsection{Semi-Analytic Method of Calculation}
We evaluate the fraction of cosmic-rays accelerated in TDE jets that undergo $p\gamma$ interactions. For $p\gamma$ interactions, we use a parameterization of the $p\gamma$ cross section that has been used in the previous publications \citep[e.g.,][]{2006PhRvD..73f3002M,2006PhRvL..97e1101M,2013PhRvL.111l1102M,2016PhRvD..93h3003S}, which utilizes the experimental cross section data and Geant4 simulation package to treat multi-pion production. However, contrary to the above previous works, we take a semi-analytical method rather than calculate neutrino spectra by solving full kinetic equations taking into account all relevant cooling processes for pions, muons, and kaons, as well as neutrino flavor mixing.  This is because such a semi-analytical method is faster and more useful for qualitative comparisons with the real data, especially when parameter scans are necessary. Note that detailed numerical treatments of pion and muon cooling can be important if the cooling is strong, as often expected in choked jets \citep{2013PhRvL.111l1102M,2016PhRvD..93h3003S}, but our semi-analytical method works reasonably well for TDE jets in which cooling effects are moderate. 

As described above, an X-ray spectral template based on observations of Sw 1644 is used to model the non-thermal spectra from jetted TDEs \citep{Burrows:2011kz}. The fraction of cosmic-rays which produce neutrinos is given by the ratio between the cooling and dynamic times $f_{p\gamma} \approx t_{\rm dyn}/t_{p\gamma}$, where $t_{\rm dyn} \approx r_{\rm em}/(\Gamma c)$. The former depends on the photohadronic cross section $\sigma_{p\gamma}(\bar{\varepsilon})$ and the TDE target photon spectra $n_{\epsilon}$ (see \S \ref{ss:photons}), and we use
\begin{equation}
\label{eq:tpgamma}
t^{-1}_{p\gamma}(\varepsilon_{p}) = \frac{1}{2}\frac{m_p^2c^3}{\varepsilon_{p}^2}\int_{\frac{\varepsilon_{\rm th}m_pc^2}{2\varepsilon_{p}}}^\infty d\varepsilon \frac{n_{\varepsilon}}{\varepsilon^2}\;\int_{\varepsilon_{\rm th}}^{\frac{2\varepsilon_{p}\varepsilon}{m_pc^2}}\,d\bar{\varepsilon} \,\bar{\varepsilon}\kappa_p(\bar{\varepsilon})\sigma_{p\gamma}(\bar{\varepsilon}),
\end{equation}
where $\bar{\varepsilon}= \varepsilon_{p}\varepsilon/m_p\,(1-\cos\theta_{p\gamma})$ is the photon energy in the CR rest frame and $\epsilon_{\rm th} \approx 140$ MeV is the threshold energy for $p\gamma$ interactions.  

\begin{figure}
\includegraphics[width=0.45\textwidth]{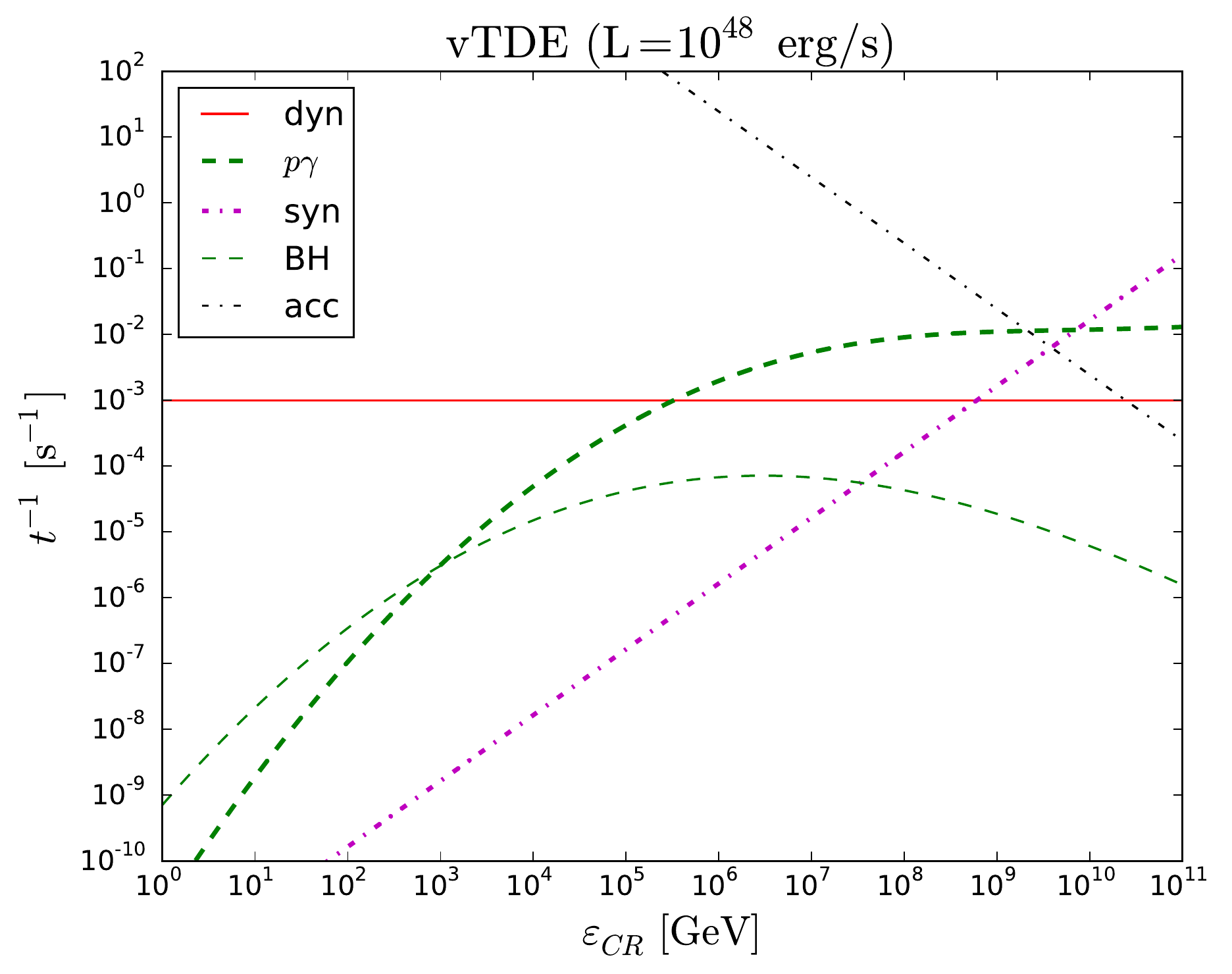}
\caption{Cooling timescales for visible TDEs as a function of parent cosmic-ray energy in the plasma rest frame of the jet. The cooling timescales were calculated in the high state, in which an isotropic X-ray luminosity $L_{[0.3,10~\rm keV]}=10^{48}\,{\rm erg/s}$ and $\xi_B=1$ are assumed, with $r_{\rm em}=3\times{10}^{14}$~cm and $\Gamma=10$. The acceleration timescale is also shown. Note that $f_{p\gamma}$ is enhanced by a factor of 10 during $t_{\rm high}$.}
\vspace{-1.\baselineskip}
\label{fig:cooling} 
\end{figure}

Our results for X-ray bright TDEs are shown in Fig. \ref{fig:cooling}. For X-ray bright TDEs, interactions with synchrotron photons are dominant.  The physical setup is similar to X-ray flares of GRBs, as pointed out by \cite{2011PhRvD..84h1301W}. For a photon spectrum given by $n_{\varepsilon}\propto \varepsilon^{-\beta}$, the efficiency is $f_{p\gamma}\propto \varepsilon_{p}^{\beta-1}$. Below the peak in $\varepsilon L_\varepsilon$, the synchrotron spectrum has a spectrum of $\beta\sim 1.5$, and it has a break at $\varepsilon_b (<\varepsilon_{\rm pk})$ (where the photon number becomes the maximum). Around the cosmic-ray energy of interest -- $\varepsilon_{p}^b\approx1.6\times10^6 (0.1~{\rm keV}/\varepsilon_b)$~GeV corresponding to the break photon energy of the TDE photon number density -- the photomeson production efficiency which is estimated to be~\citep{2006PhRvL..97e1101M}
\begin{equation}
f_{p\gamma}\sim1\frac{L_{\rm max,48}{(E_b/0.01E_{\rm pk})}^{0.5}}{r_{\rm em,14.5}\Gamma_1^2(E_{b}/1~\rm keV)}{\left(\frac{\varepsilon_{p}}{\varepsilon_{p}^b}\right)}^{\beta-1},
\end{equation}
where an additional factor of 2-3 enhancement due to multi-pion production is included. 
This is in agreement with our numerical results, taking into account the difference between a log parabolic function and power law (See Fig \ref{fig:cooling}).

\begin{figure}
\includegraphics[width=0.45\textwidth]{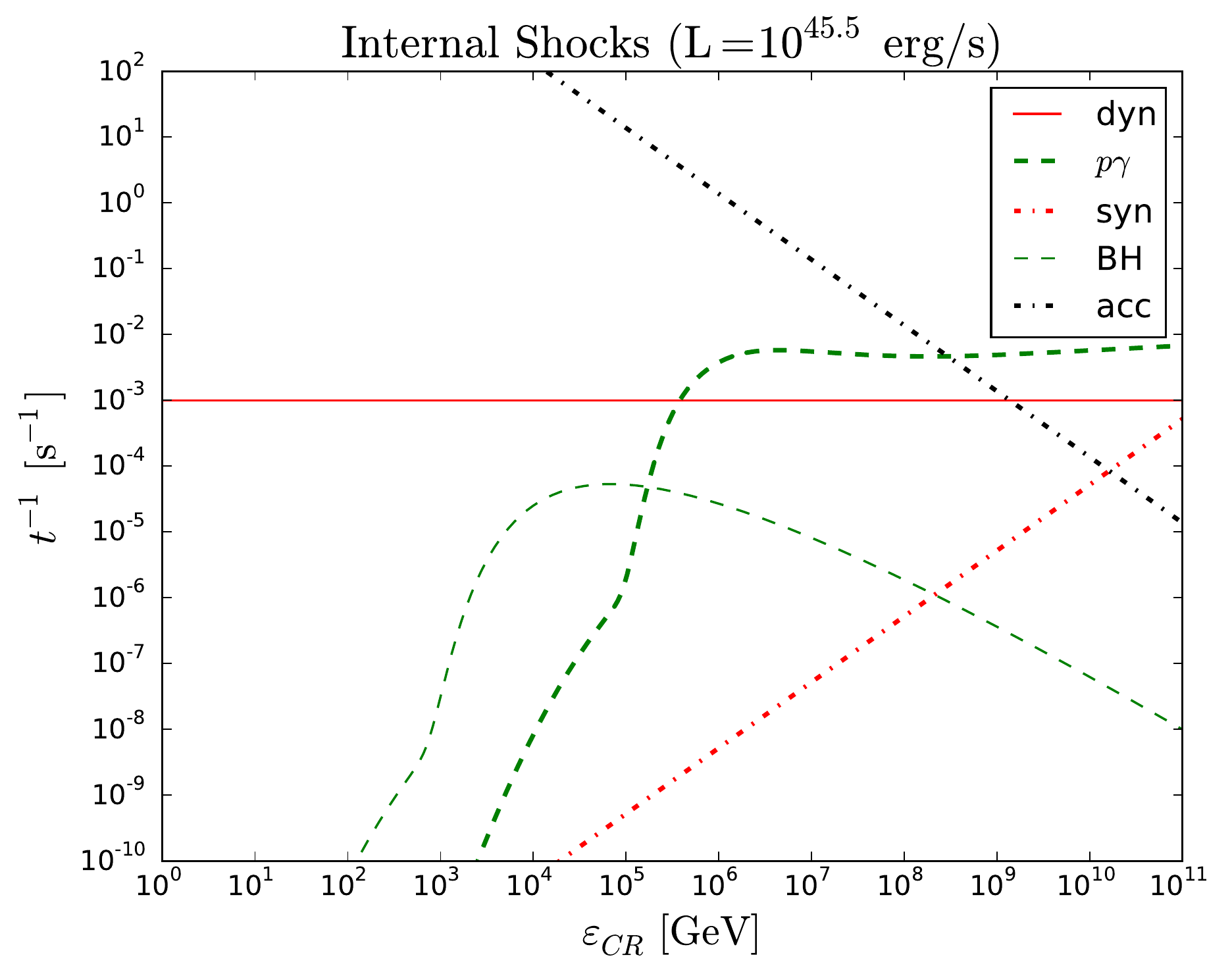}
\caption{The same as Fig.~2, but for the choked TDE jet with $L=10^{44.5}\,{\rm erg/s}$ and $\xi_B=1$ in the internal shock scenario, with $r_{\rm em}=3\times{10}^{14}$~cm and $\Gamma=10$.  The temperature of the external field is set to $T=10^5$~K, and the escape fraction is included.}
\vspace{-1.\baselineskip}
\label{fig:cooling2} 
\end{figure}
\begin{figure}
\includegraphics[width=0.45\textwidth]{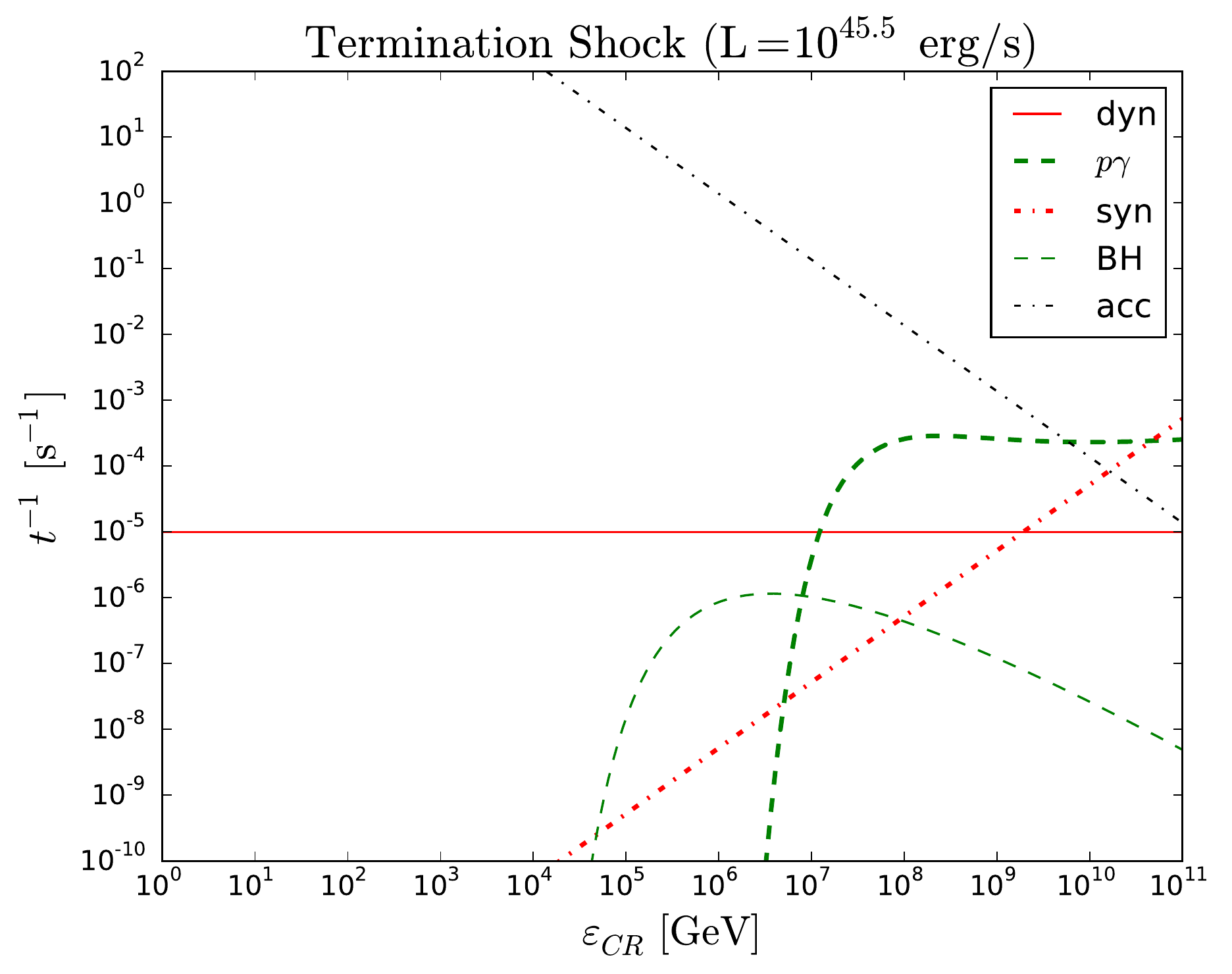}
\caption{The same as Fig.~2, but for the choked TDE jet with $L=10^{44.5}\,{\rm erg/s}$ and $\xi_B=1$ in the termination shock scenario, with $r_h=3\times{10}^{15}$~cm and $\Gamma_h=1$. The temperature of the jet head is $T_h=10^4$~K.}
\vspace{-1.\baselineskip}
\label{fig:cooling3} 
\end{figure}

Our results for TDEs with choked jets are also shown in Figs.~\ref{fig:cooling2} and \ref{fig:cooling3}, where low-power jets with $L=10^{44.5}\,{\rm erg/s}$ are assumed. The thermal radiation is relevant in both scenarios. 
In the internal shock scenario, we have 
\begin{equation}
f_{p\gamma}\sim80~\varrho_{\rm in}^{-1}r_{\rm in,13.5}^{-3}r_{14.5}^3T_5^3\Gamma_1,
\end{equation}
which is again consistent with our numerical result shown in Fig.~\ref{fig:cooling2}. Note that for the luminosity regime relevant for producing a choked jet, the non-thermal spectrum peaks at frequencies $10^{14}\,{\rm Hz} \lesssim \varepsilon/h \lesssim 3\times10^{15}\,{\rm Hz}$, leading to $p\gamma$ energy conversion at energies $\varepsilon_{p} \gtrsim 10\,{\rm PeV - 300 PeV}$. However, non-thermal contributions are smaller than thermal contributions in our case. 
Note that we do not consider $pp$ interactions. In the internal shock scenario, cosmic-rays may lose their energies via adiabatic cooling before they reach the dense material. (However, cosmic-rays accelerated at collimation shocks may experience subsequent inelastic $pp$ collisions without significant adiabatic losses~\citep{2013PhRvL.111l1102M}.) In the termination shock scenario, cosmic-rays may cause $pp$ interactions, but the $pp$ efficiency is typically small at $\sim r_h$.

Using our results on $f_{p\gamma}$, we can estimate neutrino fluxes from a single TDE event.  For a flat energy spectrum with $\varepsilon_p L_{\varepsilon_p}\propto const.$, we have
\begin{eqnarray}
\label{eq:sglevnt}
\varepsilon_\nu L_{\varepsilon_\nu} &\approx& \frac{3}{8}{\rm min}[1,f_{p\gamma}(\varepsilon_p)]f_{\rm sup}\varepsilon_p L_{\varepsilon_p}|_{\varepsilon_{p}}\nonumber\\
&\approx& \frac{3}{8}{\rm min}[1,f_{p\gamma}(\varepsilon_p)]f_{\rm sup}\frac{L_{\rm cr}}{\ln(\varepsilon_p^{\max}/\varepsilon_p^{\rm min})},
\end{eqnarray}
where $f_{\rm sup}$ is the suppression factor due to meson and muon cooling (see below). 
The observed neutrino energy is therefore related to the jet comoving proton energy for a TDE located at a redshift $z$ with bulk Lorentz factor $\Gamma$, $\varepsilon_{p} \approx 20(1+z)E_{\nu}/\Gamma$. 
The baryon loading parameter $\xi_{\rm cr} \equiv L_{\rm cr}/L_\gamma$ is the ratio of the amount of TDE jet luminosity in protons and high-energy photons \citep{2006PhRvD..73f3002M}. We assume that it has $\xi_{\rm cr}\sim\epsilon_p/\epsilon_e\sim1-100$, which is typically required for GRBs and blazars to explain UHECRs. Strictly speaking, its definition depends on the photon energy band and SED. For Sw 1644, we also define $\tilde{\xi}_{\rm cr} \equiv L_{\rm cr}/L_{[0.3,10~\rm keV]}$.

\begin{figure}
\includegraphics[width=0.45\textwidth]{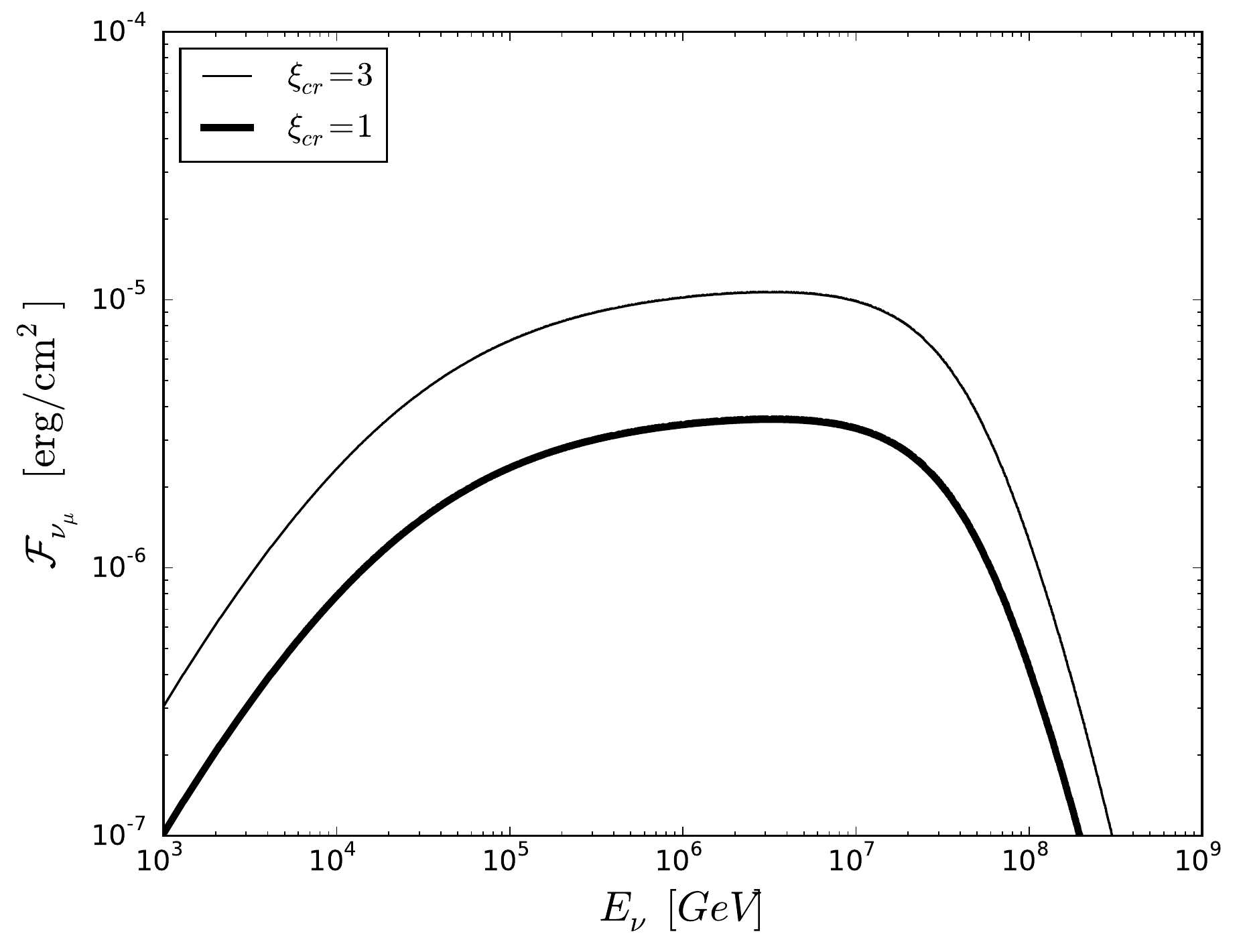}
\caption{Muon neutrino fluence for our canonical values for an X-ray bright TDE at $z\approx 0.3$. The cosmic-ray energy input is ${\mathcal E}_{\rm cr}=\xi_{\rm cr}L_\gamma t_{\rm dur}=17\times{10}^{53}~{\rm erg}$ with $\xi_{\rm cr}=1$ (thick) and $\xi_{\rm cr}=3$ (thin). Note that in our model Sw 1644 is more energetic by a factor of 2, and would have been marginally detectable with $\xi_{\rm cr}\gtrsim20$.}
\vspace{-1.\baselineskip}
\label{fig:dflux} 
\end{figure}

\begin{figure}
\includegraphics[width=0.45\textwidth]{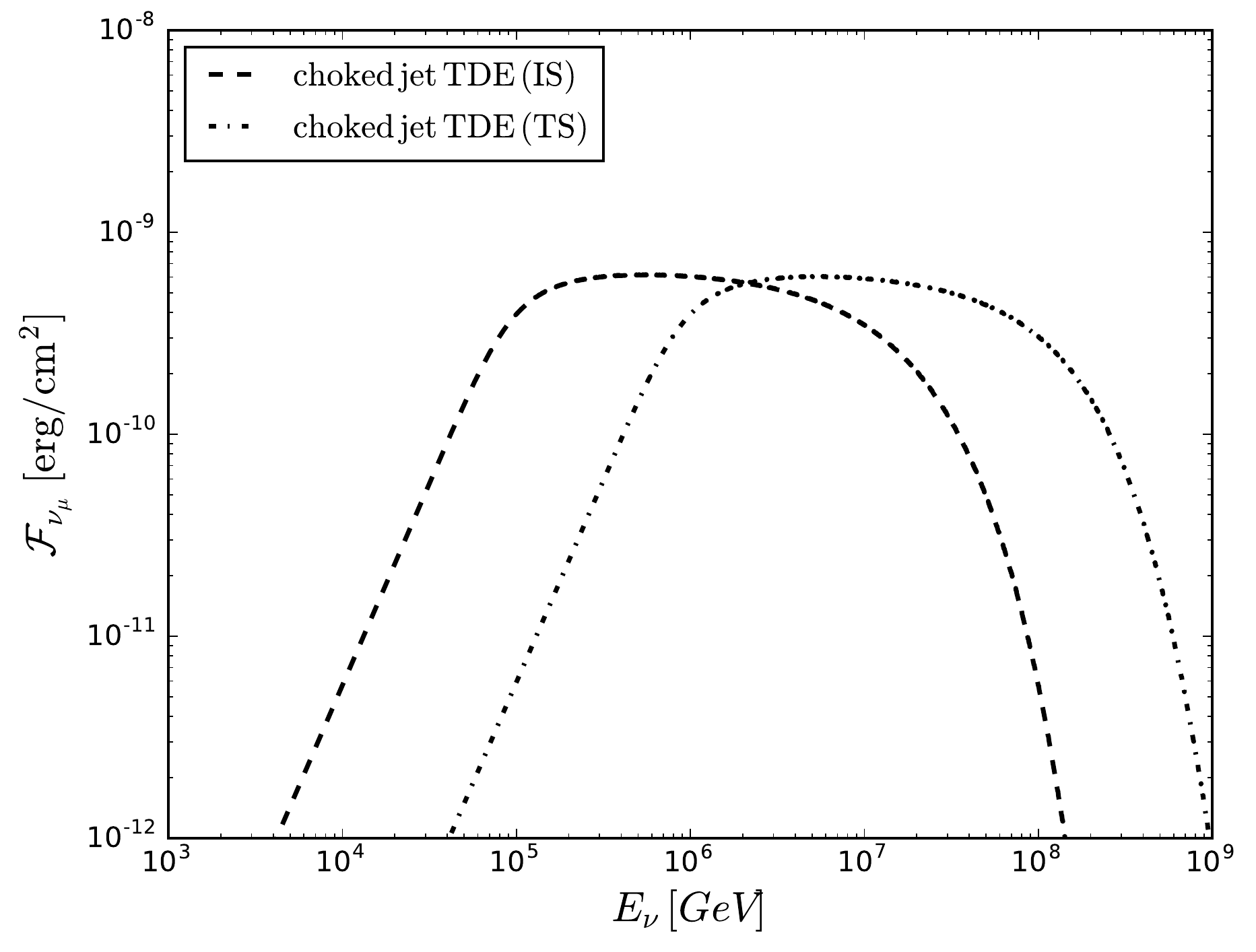}
\caption{Muon neutrino fluence for our canonical values for an X-ray dark TDE with choked jets at $z=0.3$. Neutrino contributions come from the internal shock emission region (IS) and the terminal shock region (TS), where the jet head stalls. The cosmic-ray energy input is ${\mathcal E}_{\rm cr}=\epsilon_p L t_{\rm dur}={10}^{50.5}~{\rm erg}$.}
\vspace{-1.\baselineskip}
\label{fig:dflux} 
\end{figure}

High-energy neutrino spectra can be modified when mesons and muons cool before they decay. In this work, we semi-analytically take into account the meson suppression factor, which can be approximated to be
\begin{equation}
f_{\rm sup}(\varepsilon_\pi)=\frac{t_{\rm dec-\pi}^{-1}}{t_{\rm dec-\pi}^{-1}+t_{\rm syn-\pi}^{-1}+t_{\rm IC-\pi}^{-1}+t_{\rm ad}^{-1}},
\end{equation}
where $t_{\rm dec-\pi}$ is the pion lifetime, $t_{\rm IC-\pi}$ is the pion inverse-Compton cooling time, and $t_{\rm syn}$ is the pion synchrotron cooling timescale.  For example, $t_{\rm syn-\pi}$ is given by
\begin{equation}
t^{-1}_{\rm syn-\pi}(\varepsilon_\pi) = \frac{4}{3}\sigma_{T}\frac{m_e^2}{m_\pi^4\,c^3}U_B\,\varepsilon_\pi,
\end{equation}
where $\varepsilon_\pi \approx 0.2\varepsilon_{p}$ is the pion energy in the plasma comoving frame.  Synchrotron cooling is important for pions with energy $\varepsilon_\pi \gtrsim 16\; {\rm PeV}\; \xi_B^{-1/2}\,L_{\gamma,{\rm pk},48}^{-1/2}\,r_{\rm em,14.5}\Gamma_1$. Each neutrino produced through pion decay has an average energy $\varepsilon_\nu \approx (1/4)\,\varepsilon_\pi$, meaning that the spectral break due pion synchrotron cooling occurs at observed neutrino energies $E_{\nu} \gtrsim 40$ PeV. 
\cite{Burrows:2011kz} concluded that TDE jets contain a large amount of turbulent magnetic field energy. For high magnetic fields, the charged decay products of $p\gamma$ interactions will cool before decaying if $t_{\rm syn} \lesssim t_{\rm dec}$, resulting in a cooling break in the final neutrino spectrum. 
Our results on neutrino fluences are shown in Figs. 5 and 6, and the results agree with analytical expectations based on the evaluation of $f_{p\gamma}$.

\subsection{Constraints on Neutrinos from Sw 1644 and Detection Prospects}
Among the three cases, the X-ray bright TDE jet is of particular interest since we can expect a time and space-coincidence. The three jetted TDE candidates occurred during 2011 while the IC detector was fully operational with a total of 86 strings. Furthermore, two events including Sw1644 located at $z\simeq 0.354$ occurred above Earth's northern hemisphere, and would ,therefore, be seen by IC as up going neutrinos utilizing ICs larger effective area for up going track events. Using our model of the neutrino flux from visible TDEs combined with the effective area of IC, we discuss constraints on the detectability of the jetted TDE candidate Sw 1644. 

The neutrino fluence (per flavor) for a single TDE is: 
\begin{eqnarray}
\mathcal{F}_{\nu_\mu} &\sim &0.5\times10^{-4}\;{\rm erg\,cm^{-2}}\,\left(\frac{\tilde{\xi}_{\rm cr}}{40}\right)\,\left(\frac{{\rm min}[1,f_{p\gamma}]}{0.5}\right)\,\nonumber\\
&\times&\;\left(\frac{L_{\gamma,{\rm pk}}t_{\rm hs}}{2\times10^{53}\,{\rm erg}}\right)\,\left(\frac{d_L}{2\,{\rm Gpc}}\right)^{-2}.
\end{eqnarray}
Using the IceCube effective area, we estimate that a signal neutrino could be detected from a fluence of $\mathcal{F}_{\nu_\mu}\sim 10^{-4}\,{\rm erg\,cm^{-2}}$. For Sw 1644, a naive estimate implies that $L_{0.3,10~{\rm keV}}^{\rm pk}t_{\rm high}=2\times10^{53}\,{\rm erg}$ and $\tilde{\xi}_{\rm cr}\gtrsim80$ lead to a marginally detectable event, given that $n_{\rm bkg} \approx 10^{-4}$ atmospheric background neutrino events are expected from the same region of the sky with $\tilde{\varepsilon}_{\nu_\mu}\gtrsim 10^4$ p.e.u. (Here the neutrino counts are binned in proxy energy units (p.e.u) related to the total electromagnetic energy observed by IC optical sensors.) Our results are also consistent with \cite{2011PhRvD..84h1301W}.  Comparing the calculated number with the number observed in the IC sample of up going muon neutrino events allows us to place upper limits on $\xi_{\rm cr}$. More quantitatively, one needs to take into account the effect of neutrino attenuation in the Earth. Also, the neutrino spectrum of Sw 1644 is harder than a simple $E_\nu^{-2}$ spectrum so that the expected number of muon events is smaller.

From that fact that no neutrino events were observed from the relevant part of the sky, we can set a limit of $\tilde{\xi}_{\rm cr} \lesssim 100-250$ (or $\xi_{\rm cr}\lesssim20-50$) 
With this upper limit, our model predicts that Sw 1644 would have produced a neutrino detection if it was located at redshift $z\sim0.2$ for $\xi_{\rm cr}\sim10$. The local rate of X-ray bright TDEs with peak luminosity $L_{\gamma,\rm pk} \gtrsim 10^{48}\,{\rm erg/s}$ and $L_{\gamma,\rm pk} \gtrsim 10^{47}\,{\rm erg/s}$ are $\rho_0 \sim 0.03\,{\rm Gpc^{-3}\,yr^{-1}}$ and $\rho_0 \sim 0.3\,{\rm Gpc^{-3}\,yr^{-1}}$, respectively. This implies that a TDE with that luminosity or higher occurs within $z= 0.1$ once every 10-100 years. Stacking searches can be more powerful. With X-ray sky monitors with ultimate sensitivities, which can detect TDEs up to $z\sim1$, the detection rate by IceCube could be improved to $\sim0.1-1~{\rm yr}^{-1}$.

\section{Diffuse Neutrino Intensity} 
\label{sec:calc_flux}

We calculate the diffuse flux from jetted TDEs. 
The Eddington luminosity of a $10^6 M_\odot$ SMBH is $L_{\rm Edd,\rm BH}\sim 10^{44} M_{\rm BH,6}\,{\rm erg\,s^{-1}}$. The corresponding isotropic equivalent jet luminosity is $L \sim 2\times10^{46}\theta_{j,-1}^{-2}\,{\rm erg\,s^{-1}}$. One can calculate the diffuse neutrino flux from ``visible" (i.e., not choked) jetted TDEs (e.g., Sw 1644) by integrating luminosities for $10^{44.5}\,{\rm erg\, s^{-1}} < L_\gamma < 10^{49.5}\,{\rm erg\, s^{-1}}$.  In this work, we allow $L_\gamma < 10^{46}\,{\rm erg\, s^{-1}}$ to consider a maximum contribution, but this does not change our results for the luminosity function we use. 
We consider choked jet contributions by extrapolating the luminosity function down to $L_\gamma < 10^{44.5}\,{\rm erg\, s^{-1}}$, which can be choked by a $0.5M_\odot$ material with a radius of $r_{\rm out} \sim 3\times10^{15}\,{\rm cm}$ (See $\S$ IIA). 

One of the important predictions for TDEs is that they typically have a negative or weak evolution.  
We use the following TDE redshift evolution function \citep{2015ApJ...812...33S},
\begin{equation}
\label{eq:fz}
f_{\rm TDE}(z) = \left[(1+z)^{0.2\eta}+\left(\frac{1+z}{1.43}\right)^{-3.2\eta} + \left(\frac{1+z}{2.66}\right)^{-7.0\eta}\right]^{\frac{1}{\eta}}
\end{equation}
with $\eta = -2$. Note that this redshift distribution peaks for $z \approx 0$, whereas the usual star-formation rate and GRB rate tend to peak around $z\sim1-3$, leading to larger values of $\xi_z$ . 

The individual neutrino fluxes are then integrated over redshift and isotropic equivalent luminosity \citep{2014PhRvD..90b3007M} 
\begin{eqnarray}
\label{eq:difflux} 
\Phi_\nu &=& \frac{c}{4\pi H_0}\int_{z_{\rm min}}^{z_{\rm max}}\,dz \,\int_{L_{\rm min}}^{L_{\rm max}}\,dL_{\gamma}\nonumber\\
&\times&\frac{d\rho_{\rm TDE}(z,L_\gamma)/dL_\gamma}{\sqrt{\Omega_M\,(1+z)^3 + \Omega_\Lambda}}\,\frac{L_{E'_\nu}(L_\gamma)}{E'_\nu}t_{\rm eng},
\end{eqnarray}
where $\frac{d\rho_{\rm TDE}}{dL_\gamma} = \rho_{0} f(z)\Lambda_{TDE}(L_\gamma)$. Although observational uncertainties are large at present, both conventional (i.e. non-jetted) and Swift X-ray TDEs appear to share a luminosity function over seven orders of magnitude in observed peak luminosity with 
\begin{equation}
\label{eq:Lfn}
\Lambda_{\rm TDE}(L_{\gamma}) \propto \left(\frac{L_{\gamma,\rm pk}}{L_{m,\rm pk}}\right)^{-\alpha},
\end{equation}
$L_{m,\rm pk} = 10^{48}\,{\rm erg~s^{-1}}$, and $\alpha = 2.0\pm 0.05$ \citep{2015ApJ...812...33S}. Eq. \ref{eq:Lfn} is derived using the {\it peak} X-ray luminosity of an event, whereas we take $L_\gamma$ to be the average luminosity over the initial $\sim 10^6$ s of a jetted TDE. Again using observations of SW 1644, we assume the rough relationship $L_{\gamma,\rm pk}\simeq12\, L_\gamma$ (compare the maximum and median X-ray luminosity of SW 1644 in Supplementary Table 7 of \cite{Burrows:2011kz}). 

\begin{figure}
\includegraphics[width=0.45\textwidth]{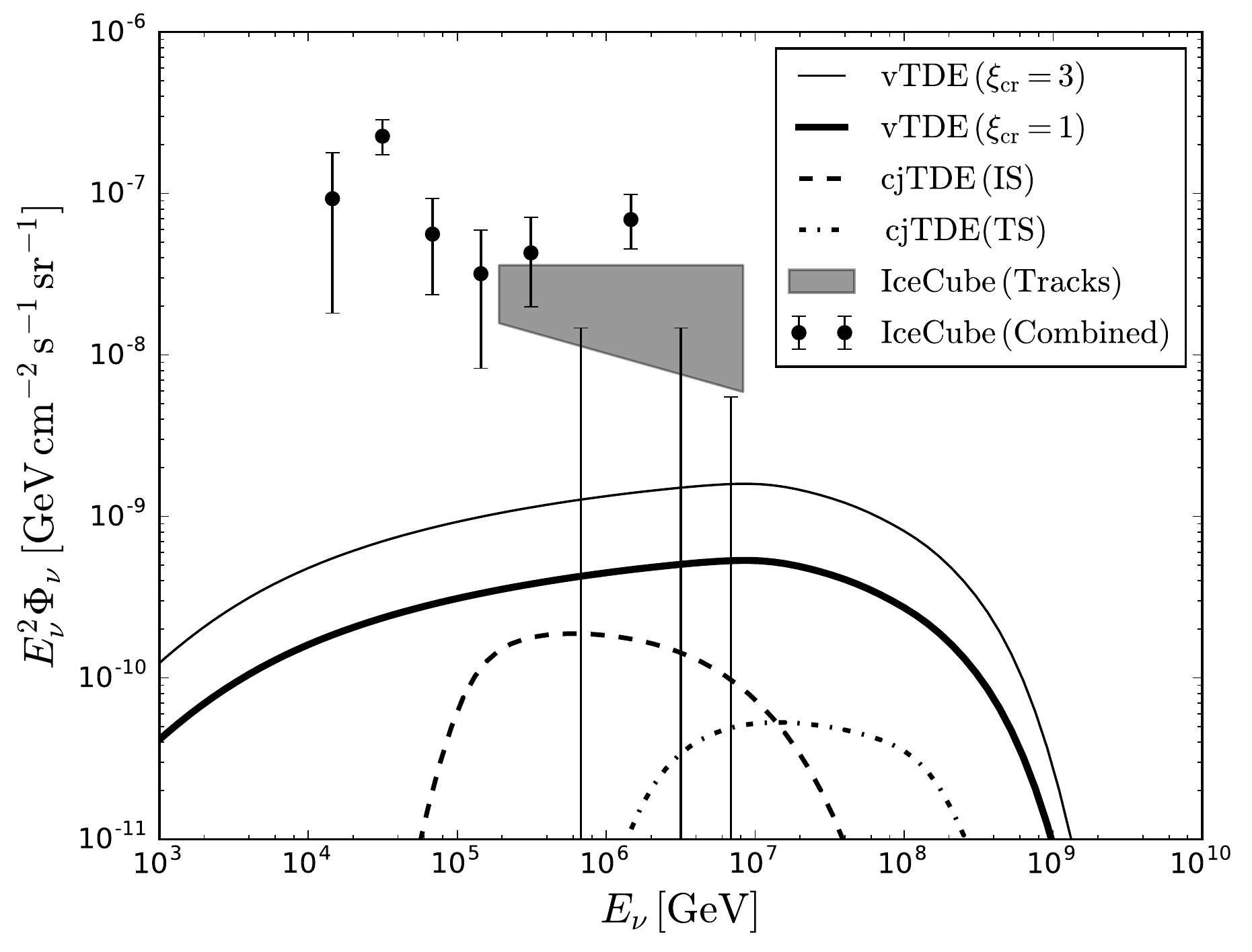}
\caption{
Contributions to the diffuse neutrino background due to $p\gamma$ interactions from X-ray bright visible jets and possible choked jets. For successful jets leading to X-ray bright TDEs, the cosmic-ray luminosity is given by $\xi_{\rm cr}=L_{\rm cr}/L_\gamma=1-3$. For choked jets, the internal shock and termination shock scenarios are considered, and the cosmic-ray luminosity is assumed to be comparable to the total luminosity $L$.   
For the diffuse neutrino data, we use the muon neutrino data obtained by IceCube with the multiplication by a factor of 3 \citep{IceCubeCollaboration:2016ta}. 
}
\vspace{-1.\baselineskip}
\label{fig:dflux} 
\end{figure}

Using $t_{p\gamma}(\varepsilon_{p},L_{\gamma,\rm pk})$ (see Eq. \ref{eq:tpgamma}), we integrate Eq. \ref{eq:difflux} and show the results of the diffuse neutrino flux from visible and choked-jet TDEs in Fig. \ref{fig:dflux}. The results can be understood by virtue of analytical estimates. The all flavor diffuse flux of neutrinos from extragalactically distributed sources can be estimated from the amount of cosmic-ray energy released per burst ${\mathcal E}_{\rm cr}=\xi_{\rm cr}{\mathcal E}_\gamma$ and the local rate of such explosions $\rho_{0}$ as
\begin{eqnarray}
E_\nu^2\Phi_{\nu}&\approx& \frac{3}{8}\frac{c}{4\pi}t_H\xi_z \,{\rm min}[1,f_{p\gamma}]f_{\rm sup}f_{\rm cho}\frac{{\mathcal E}_{\rm cr}}{\mathcal C}\rho_0\nonumber\\
&\sim&1\times{10}^{-9}~{\rm GeV}~{\rm cm}^{-2}~{\rm s}^{-1}~{\rm sr}^{-1}~{\rm min}[1,f_{p\gamma}]f_{\rm sup}\nonumber\\
&\times&f_{\rm cho}\xi_{\rm cr,0.5}{\mathcal E}_{\gamma,53.8}(\rho_0/0.1~{\rm Gpc^{-3}}~{\rm yr}^{-1})(\xi_z/0.5)
\,\,\,\,\,\,\,\,\,\,\,\,\,\,\,\,\,\,
\end{eqnarray}
where $f_{\rm cho}(\geq1)$ is the enhancement factor (for energy budgets) due to the existence of choked jets \citep{2013PhRvL.111l1102M}. 
We assume the Hubble time $t_H\sim 13.7$ Gyrs, a red-shift correction factor $\xi_z\lesssim0.6$ since most TDE are located at low redshift (See Eq. \ref{eq:fz}). The amount of cosmic-ray energy released from a TDE, approximated by $ \varepsilon_{p}Q_{\varepsilon_{p}}\sim t_{\rm eng}L_{\rm cr}/\mathcal{C}$. The factor $\mathcal{C} = \ln(\varepsilon_{p,\rm max}/\varepsilon_{p,\rm min})\sim20$ converts the total bolometric cosmic-ray energy into the cosmic-ray energy per logarithmic energy interval which is generally what is expressed for the diffuse neutrino flux. 
The rate of TDEs is expressed in terms of the local TDE rate $\rho_0 = 0.03^{+0.04}_{-0.02}\,{\rm Gpc^{-3}\,yr^{-1}}$ for events with peak luminosity greater than $L_m = 10^{48}\,{\rm erg/s}$, and a luminosity function $\rho_0f(z)\Lambda_{\rm TDE}(L_\gamma)\equiv \frac{\rho_{\rm TDE}}{dL_{\gamma}}\propto L_\gamma^{-\alpha}$ (See Eq. \ref{eq:Lfn}). 
As a result, the TDE rates above $L_\gamma\sim10^{46}\,{\rm erg/s}$ and  $L_\gamma\sim10^{46.5}\,{\rm erg/s}$ are $\rho_0 \sim 0.3\,{\rm Gpc^{-3}\,yr^{-1}}$ and $\rho_0 \sim 0.1\,{\rm Gpc^{-3}\,yr^{-1}}$, respectively. 

For our luminosity function with $\alpha=2$, the energy generation rate of visible jets and choked jets are similar. Thus, the maximum enhancement factor is $f_{\rm cho}$ cannot be as large as $\sim10-1000$. Thus, the neutrino flux from choked jets is comparable to that of visible jets and IceCube's neutrino flux cannot be explained. However, the luminosity function is uncertain. If the luminosity function is significantly steeper than the one we use and almost all TDEs have choked jets, $f_{\rm cho}$ could be as large as $\sim100-1000$. With such a bit extreme assumption, it is possible for them to significantly contribute the diffuse neutrino flux.  

The calculated fluxes are significant, but below the observed diffuse flux of IC neutrinos \citep{IceCubeCollaboration:2016ta}, even with optimistic parameters. We have other constraints from multiplet searches. Using the absence of significant clustering in the 6 or 7 year data of IceCube, \cite{Murase:2016ez} obtained 
\begin{equation}\label{eq:n0}
n_0^{\rm eff}\gtrsim1.1\times10^{-7}\,{\rm Mpc^{-3}}\,q_L^2\left(\frac{\xi_z}{3}\right)^{-3}F^{-3}_{\rm lim, -9}\left(\frac{\Delta \Omega}{2\pi}\right)^2.
\end{equation}
By replacing $n_0$ with $\rho_0 t_{\rm dur}^3/T_{\rm IC}^2$, the above result is readily rewritten as
\begin{equation}\label{eq:rho0}
\rho_0^{\rm eff}\gtrsim1.7\times10^{3}\,{\rm Gpc^{-3}}~{\rm yr^{-1}}\,\frac{q_L^2{({\Delta \Omega/2\pi)}^2{(T_{\rm IC}/6~{\rm yr})}^2}}{\xi_z^3F^{3}_{\rm lim, -6.9}t_{\rm dur,6}^3}.
\end{equation}
For TDEs with $t_{\rm eng}\sim{10}^{6}$~s, we have used the sensitivity $F_{\rm lim}\sim{10}^{-7}~{\rm GeV}~{\rm cm}^{-2}~{\rm s}^{-1}$. Note that the above limit is weaker by a factor for hard spectra \citep{Murase:2016ez}, but we can conclude that it is very unlikely that X-ray bright TDEs are the sources of IceCube's neutrinos.  X-ray bright TDEs with the local rate of visible TDEs $\rho_0 \sim 0.1\,{\rm Gpc^{-3}\,yr^{-1}}$ can give $\lesssim5-10$\% of the diffuse neutrino flux, which is consistent with our diffuse flux calculations. In other words, $\xi_{\rm cr}\gtrsim100$ is unlikely. 
However, in principle, X-ray dim TDEs with choked jets can avoid this constraint by achieving $f_{\rm cho}\gg1$~\footnote{For choked ultra-long (or low-luminosity) GRB jets with $\alpha=2.3$, $f_{\rm cho}\gtrsim2-3$ (or $f_{\rm cho}\gtrsim1$) can avoid constraints thanks to larger $\xi_z$ as long as cosmic-rays can be accelerated.}. The luminosity function should be different from that obtained by \cite{2015ApJ...812...33S}; that is, $\alpha\gtrsim3.2-4.0$ is required. 

X-ray bright TDEs are also expected to be $\gamma$-ray dim. While GeV-TeV $\gamma$-rays are expected from $p\gamma$ interactions, they are attenuated by the X-rays that are the target photons for cosmic-ray interactions. The $\gamma$-ray optical depth in X-ray bright sources can be related to the fraction of cosmic-rays that undergo $p\gamma$ interactions, $\tau_{\gamma\gamma} \sim \sigma_{\gamma\gamma}/(\kappa_p\sigma_{p\gamma})\,f_{p\gamma} \sim 1000\,f_{p\gamma}$ \citep{Murase:2016ck}. From Fig. \ref{fig:cooling} we see that $f_{p\gamma}\sim0.1$ for $\varepsilon_{p}\gtrsim 10$ TeV. Since $\varepsilon/\varepsilon_p \sim m_e^2c^2/(0.15~{\rm GeV}~m_p)$, X-ray bright TDEs are opaque to $\gamma$-rays with energies $E_\gamma \gtrsim 10-100$~MeV, which is consistent with \cite{Burrows:2011kz}.

\section{Conclusion and Discussion}

We studied different possibilities of high-energy neutrino production inside TDE jets. 
Sw 1644 is the closest and brightest jetted TDE candidate, and the non-detection would give us a limit on the cosmic-ray loading factor, $\xi_{\rm cr}\lesssim20-50$. 
If Sw 1644 had occurred within $z \sim 0.1-0.2$, it would have produced a definitive neutrino detection. Since such a bright TDE only occurs once every 10-100 years within that distance, the association of X-ray bright TDEs with neutrinos will require aggregate neutrino coincidence searches such as stacking analyses. If we have X-ray all-sky monitors with ultimate sensitivities allowing us to see TDEs up to $z\sim1$, IceCube may detect neutrino signals from TDEs, with $0.1-1~{\rm yr}^{-1}$.  But new X-ray satellites with a wide field of view and a sensitivity of $F_{\rm lim}\sim5\times{10}^{-11}~{\rm erg}~{\rm cm}^{-2}~{\rm s}^{-1}~L_{\gamma,\rm pk,48}{(d_L/6.6~\rm Gpc)}^{-2}$ would be necessary for efficient searches of high-energy neutrinos from jetted TDEs. 
With IceCube-Gen2~\citep{2014arXiv1412.5106I}, our results imply that it is possible to detect the brightest TDE events such as Sw 1644. 

The method we use to determine the limit on the neutrino brightness of Sw 1644 can be applied to other non-$\gamma$-ray bright sources (e.g., Type Ibc SNe), assuming that a consistent model of the broadband electromagnetic and neutrino emission is formulated. It is also important to determine the duration of the neutrino emission, and any time delays between photon and neutrino arrival times since this affects the background rate of IC neutrinos. Understanding the size of the signal time interval also affects the energy limit in the search. For example, using one year of IC data required a cut at $\tilde{\varepsilon}_{\nu_\mu} \gtrsim 10^4$ p.e.u, while a time window of $10^6$ s around the triggering time of Sw 1644 would lower the atmospheric background rate of neutrinos and lead to a lower energy threshold of $\tilde{\varepsilon}_{\nu_\mu} \gtrsim 10^3$ p.e.u. Future searches of $\gamma$-ray dim transient sources (i.e., sources other than GRBs and Blazar flares) will require an optimization of the threshold energy and time interval search size. 

We also found that the flux and spectral index of the diffuse neutrino flux from electromagnetically bright TDEs are typically sub-dominant as an origin of IceCube's neutrino flux, but could be significant at very high energies ($E_\nu \gtrsim 1$ PeV).
As in $\gamma$-ray bursts \citep{2015NatCo...6E6783B} and X-ray flares in GRB afterglows \citep{2006PhRvL..97e1101M}, neutrinos flares associated with TDEs, which may have ${10}^{-9}~{\rm GeV}~{\rm cm}^{-2}~{\rm s}^{-1}~{\rm sr}^{-1}$, are interesting targets for next-generation neutrino telescopes that are suitable for extremely high-energies, such as IceCube-Gen2~\citep{2014arXiv1412.5106I}, Askaryan Radio Array~\citep{2012APh....35..457A}, ExaVolt Antenna~\citep{2011APh....35..242G}, and Giant Radio Array for Neutrino Detection~\citep{2016EPJWC.11603005M}.

We discuss possible contributions made by TDEs with choked jets. Although this scenario is still speculative, we found that the magnitude of the diffuse neutrino flux expected from choked jet TDEs is typically low to be compatible with IC measurements. However, we caution that our estimate is based on the large extrapolation of the luminosity function. The luminosity function obtained by \cite{2015ApJ...812...33S}, $\Lambda_{\rm TDE}(L_\gamma)\propto L_\gamma^{-2}$, all TDEs with different luminosities contribute approximately the same amount to the diffuse cosmic-ray flux. The neutrino flux is dominated by TDEs with $f_{p\gamma}\gtrsim1$. 
The situation is different from other classes of choked jet objects \citep[e.g.,][for low-power GRBs]{2013PhRvL.111l1102M,2016PhRvD..93h3003S} that have been found to be more efficient neutrino producers. For example, low-luminosity GRBs have $\Lambda_{\rm LL GRB} \propto L_\gamma^{-2.3\pm0.2}$ so that the lowest-luminosity events largely contribute to the diffuse neutrino flux. 
Because of large uncertainty, it would be possible to assume steeper values of $\alpha$, and then one could increase the contributions from choked jets. A steep luminosity function leading to $f_{\rm cho}\sim100-1000$ would be required to explain the IceCube data. Besides this issue, the setup expressed by Eq.~(\ref{eq:density}) should be tested by observations. For example, the radio emission is useful to probe densities of the circumnuclear material around a SMBH \citep{2017MNRAS.464.2481G}. Also, how sub-Eddington luminosity jets are launched by spinning BHs should be justified.  


\medskip
\acknowledgments
We thank Xiang-Yu Wang for discussion and useful suggestions. While we were finalizing this manuscript, we became aware of the independent work by Dai \& Fang (2016). We also thank Jane Da, Ke Fang, and Nick Stone for helpful communications.
The work of K.M. is supported by NSF Grant No. PHY-1620777.  N.S. and P.M. acknowledge partial support from NASA NNX13AH50G.
\bibliographystyle{apj}
\bibliography{tde16}

\begin{thebibliography}{}
\expandafter\ifx\csname natexlab\endcsname\relax\def\natexlab#1{#1}\fi

\bibitem[{{Aartsen} {et~al.}(2013){Aartsen}, {Abbasi}, {Abdou}, {Ackermann},
  {Adams}, {Aguilar}, {Ahlers}, {Altmann}, {Auffenberg}, {Bai}, \&
  et~al.}]{ICdiscovery}
{Aartsen}, M.~G., {Abbasi}, R., {Abdou}, Y., {et~al.} 2013, Physical Review
  Letters, 111, 021103

\bibitem[{{Aartsen} {et~al.}(2014){Aartsen}, {Ackermann}, {Adams}, {Aguilar},
  {Ahlers}, {Ahrens}, {Altmann}, {Anderson}, \& et~al.}]{2014arXiv1412.5106I}
{Aartsen}, M.~G., {Ackermann}, M., {Adams}, J., {et~al.} 2014, ArXiv e-prints,
  arXiv:1412.5106

\bibitem[{Aartsen {et~al.}(2015{\natexlab{a}})Aartsen, Abraham, Ackermann,
  Adams, Aguilar, Ahlers, Ahrens, Altmann, Anderson, Archinger, Arguelles,
  Arlen, Auffenberg, Bai, Barwick, Baum, Bay, Beatty, Becker~Tjus, Becker,
  Beiser, BenZvi, Berghaus, Berley, Bernardini, Bernhard, Besson, Binder,
  Bindig, Bissok, Blaufuss, Blumenthal, Boersma, Bohm, B{\"o}rner, Bos, Bose,
  B{\"o}ser, Botner, Braun, Brayeur, Bretz, Brown, Buzinsky, Casey, Casier,
  Cheung, Chirkin, Christov, Christy, Clark, Classen, Coenders, Cowen,
  Cruz~Silva, Daughhetee, Davis, Day, de~Andr{\'e}, De~Clercq, Dembinski,
  De~Ridder, Desiati, de~Vries, de~Wasseige, de~With, DeYoung,
  D{\'\i}az-V{\'e}lez, Dumm, Dunkman, Eagan, Eberhardt, Ehrhardt, Eichmann,
  Euler, Evenson, Fadiran, Fahey, Fazely, Fedynitch, Feintzeig, Felde,
  Filimonov, Finley, Fischer-Wasels, Flis, Fuchs, Gaisser, Gaior, Gallagher,
  Gerhardt, Ghorbani, Gier, Gladstone, Glagla, Gl{\"u}senkamp, Goldschmidt,
  Golup, Gonzalez, Goodman, G{\'o}ra, Grant, Gretskov, Groh, Gro{\ss}, Ha,
  Haack, Haj~Ismail, Hallgren, Halzen, Hansmann, Hanson, Hebecker, Heereman,
  Helbing, Hellauer, Hellwig, Hickford, Hignight, Hill, Hoffman, Hoffmann,
  Holzapfel, Homeier, Hoshina, Huang, Huber, Huelsnitz, Hulth, Hultqvist, In,
  Ishihara, Jacobi, Japaridze, Jero, Jurkovic, Kaminsky, Kappes, Karg, Karle,
  Kauer, Keivani, Kelley, Kemp, Kheirandish, Kiryluk, Kl{\"a}s, Klein, Kohnen,
  Kolanoski, Konietz, Koob, K{\"o}pke, Kopper, Kopper, Koskinen, Kowalski,
  Krings, Kroll, Kroll, Kunnen, Kurahashi, Kuwabara, Labare, Lanfranchi,
  Larson, Lesiak-Bzdak, Leuermann, Leuner, L{\"u}nemann, Madsen, Maggi, Mahn,
  Maruyama, Mase, Matis, Maunu, McNally, Meagher, Medici, Meli, Menne, Merino,
  Meures, Miarecki, Middell, Middlemas, Miller, Mohrmann, Montaruli, Morse,
  Nahnhauer, Naumann, Niederhausen, Nowicki, Nygren, Obertacke, Olivas,
  Omairat, O'Murchadha, Palczewski, Paul, Pepper, P{\'e}rez de~los Heros,
  Pfendner, Pieloth, Pinat, Posselt, Price, Przybylski, P{\"u}tz, Quinnan,
  R{\"a}del, Rameez, Rawlins, Redl, Reimann, Relich, Resconi, Rhode, Richman,
  Richter, Riedel, Robertson, Rongen, Rott, Ruhe, Ruzybayev, Ryckbosch, Saba,
  Sabbatini, Sander, Sandrock, Sandroos, Sarkar, Schatto, Scheriau, Schimp,
  Schmidt, Schmitz, Schoenen, Sch{\"o}neberg, Sch{\"o}nwald, Schukraft,
  Schulte, Seckel, Seunarine, Shanidze, Smith, Soldin, Spiczak, Spiering,
  Stahlberg, Stamatikos, Stanev, Stanisha, Stasik, Stezelberger, Stokstad,
  St{\"o}{\ss}l, Strahler, Str{\"o}m, Strotjohann, Sullivan, Sutherland,
  Taavola, Taboada, Ter-Antonyan, Terliuk, Te{\v s}i{\'c}, Tilav, Toale, Tobin,
  Tosi, Tselengidou, Unger, Usner, Vallecorsa, Vandenbroucke, van Eijndhoven,
  Vanheule, van Santen, Veenkamp, Vehring, Voge, Vraeghe, Walck, Wallace,
  Wallraff, Wandkowsky, Weaver, Wendt, Westerhoff, Whelan, Whitehorn, Wichary,
  Wiebe, Wiebusch, Wille, Williams, Wissing, Wolf, Wood, Woschnagg, \&
  X...}]{2015ApJ...809...98A}
Aartsen, M.~G., Abraham, K., Ackermann, M., {et~al.} 2015{\natexlab{a}}, ApJ,
  809, 98

\bibitem[{Aartsen {et~al.}(2015{\natexlab{b}})Aartsen, Abraham, Ackermann,
  Adams, Aguilar, Ahlers, Ahrens, Altmann, Anderson, Archinger, Arguelles,
  Arlen, Auffenberg, Bai, Barwick, Baum, Bay, Beatty, Tjus, Becker, Beiser,
  BenZvi, Berghaus, Berley, Bernardini, Bernhard, Besson, Binder, Bindig,
  Bissok, Blaufuss, Blumenthal, Boersma, Bohm, B{\"o}rner, Bos, Bose,
  B{\"o}ser, Botner, Braun, Brayeur, Bretz, Brown, Buzinsky, Casey, Casier,
  Cheung, Chirkin, Christov, Christy, Clark, Classen, Coenders, Cowen, Silva,
  Daughhetee, Davis, Day, de~Andr{\'e}, De~Clercq, Dembinski, De~Ridder,
  Desiati, de~Vries, de~Wasseige, de~With, DeYoung, D{\'\i}az-V{\'e}lez, Dumm,
  Dunkman, Eagan, Eberhardt, Ehrhardt, Eichmann, Euler, Evenson, Fadiran,
  Fahey, Fazely, Fedynitch, Feintzeig, Felde, Filimonov, Finley,
  Fischer-Wasels, Flis, Fuchs, Glagla, Gaisser, Gaior, Gallagher, Gerhardt,
  Ghorbani, Gier, Gladstone, Gl{\"u}senkamp, Goldschmidt, Golup, Gonzalez,
  Goodman, G{\'o}ra, Grant, Gretskov, Groh, Gro{\ss}, Ha, Haack, Ismail,
  Hallgren, Halzen, Hansmann, Hanson, Hebecker, Heereman, Helbing, Hellauer,
  Hellwig, Hickford, Hignight, Hill, Hoffman, Hoffmann, Holzapfe, Homeier,
  Hoshina, Huang, Huber, Huelsnitz, Hulth, Hultqvist, In, Ishihara, Jacobi,
  Japaridze, Jero, Jurkovic, Kaminsky, Kappes, Karg, Karle, Kauer, Keivani,
  Kelley, Kemp, Kheirandish, Kiryluk, Kl{\"a}s, Klein, Kohnen, Kolanoski,
  Konietz, Koob, K{\"o}pke, Kopper, Kopper, Koskinen, Kowalski, Krings, Kroll,
  Kroll, Kunnen, Kurahashi, Kuwabara, Labare, Lanfranchi, Larson, Lesiak-Bzdak,
  Leuermann, Leuner, L{\"u}nemann, Madsen, Maggi, Mahn, Maruyama, Mase, Matis,
  Maunu, McNally, Meagher, Medici, Meli, Menne, Merino, Meures, Miarecki,
  Middell, Middlemas, Miller, Mohrmann, Montaruli, Morse, Nahnhauer, Naumann,
  Niederhausen, Nowicki, Nygren, Obertacke, Olivas, Omairat, O'Murchadha,
  Palczewski, Paul, Pepper, de~los Heros, Pfendner, Pieloth, Pinat, Posselt,
  Price, Przybylski, P{\"u}tz, Quinnan, R{\"a}del, Rameez, Rawlins, Redl,
  Reimann, Relich, Resconi, Rhode, Richman, Richter, Riedel, Robertson, Rongen,
  Rott, Ruhe, Ruzybayev, Ryckbosch, Saba, Sabbatini, Sander, Sandrock,
  Sandroos, Sarkar, Schatto, Scheriau, Schimp, Schmidt, Schmitz, Schoenen,
  Sch{\"o}neberg, Sch{\"o}nwald, Schukraft, Schulte, Seckel, Seunarine,
  Shanidze, Smith, Soldin, Spiczak, Spiering, Stahlberg, Stamatikos, Stanev,
  Stanisha, Stasik, Stezelberger, Stokstad, St{\"o}{\ss}l, Strahler, Str{\"o}m,
  Strotjohann, Sullivan, Sutherland, Taavola, Taboada, Ter-Antonyan, Terliuk,
  Te{\v s}i{\'c}, Tilav, Toale, Tobin, Tosi, Tselengidou, Unger, Usner,
  Vallecorsa, van Eijndhoven, Vandenbroucke, van Santen, Vanheule, Veenkamp,
  Vehring, Voge, Vraeghe, Walck, Wallraff, Wandkowsky, Weaver, Wendt,
  Westerhoff, Whelan, Whitehorn, Wichary, Wiebe, Wiebusch, Wille, Williams,
  Wissing, Wolf, Wood, Woschnagg, Xu, \& Xu}]{Aartsen:2015de}
---. 2015{\natexlab{b}}, Phys. Rev. Lett., 115, 081102

\bibitem[{{Ackermann} {et~al.}(2012){Ackermann}, {Ajello}, {Allafort},
  {Schady}, {Baldini}, {Ballet}, {Barbiellini}, {Bastieri}, {Bellazzini},
  {Blandford}, {Bloom}, {Borgland}, {Bottacini}, {Bouvier}, {Bregeon},
  {Brigida}, {Bruel}, {Buehler}, {Buson}, {Caliandro}, {Cameron}, {Caraveo},
  {Cavazzuti}, {Cecchi}, {Charles}, {Chaves}, {Chekhtman}, {Cheung}, {Chiang},
  {Chiaro}, {Ciprini}, {Claus}, {Cohen-Tanugi}, {Conrad}, {Cutini},
  {D'Ammando}, {de Palma}, {Dermer}, {Digel}, {do Couto e Silva},
  {Dom{\'{\i}}nguez}, {Drell}, {Drlica-Wagner}, {Favuzzi}, {Fegan}, {Focke},
  {Franckowiak}, {Fukazawa}, {Funk}, {Fusco}, {Gargano}, {Gasparrini},
  {Gehrels}, {Germani}, {Giglietto}, {Giordano}, {Giroletti}, {Glanzman},
  {Godfrey}, {Grenier}, {Grove}, {Guiriec}, {Gustafsson}, {Hadasch},
  {Hayashida}, {Hays}, {Jackson}, {Jogler}, {Kataoka}, {Kn{\"o}dlseder},
  {Kuss}, {Lande}, {Larsson}, {Latronico}, {Longo}, {Loparco}, {Lovellette},
  {Lubrano}, {Mazziotta}, {McEnery}, {Mehault}, {Michelson}, {Mizuno}, {Monte},
  {Monzani}, {Morselli}, {Moskalenko}, {Murgia}, {Tramacere}, {Nuss},
  {Greiner}, {Ohno}, {Ohsugi}, {Omodei}, {Orienti}, {Orlando}, {Ormes},
  {Paneque}, {Perkins}, {Pesce-Rollins}, {Piron}, {Pivato}, {Porter},
  {Rain{\`o}}, {Rando}, {Razzano}, {Razzaque}, {Reimer}, {Reimer}, {Reyes},
  {Ritz}, {Rau}, {Romoli}, {Roth}, {S{\'a}nchez-Conde}, {Sanchez}, {Scargle},
  {Sgr{\`o}}, {Siskind}, {Spandre}, {Spinelli}, {Stawarz}, {Suson},
  {Takahashi}, {Tanaka}, {Thayer}, {Thompson}, {Tibaldo}, {Tinivella},
  {Torres}, {Tosti}, {Troja}, {Usher}, {Vandenbroucke}, {Vasileiou},
  {Vianello}, {Vitale}, {Waite}, {Winer}, {Wood}, \&
  {Wood}}]{2012Sci...338.1190A}
{Ackermann}, M., {Ajello}, M., {Allafort}, A., {et~al.} 2012, Science, 338,
  1190

\bibitem[{{Allison} {et~al.}(2012){Allison}, {Auffenberg}, {Bard}, {Beatty},
  {Besson}, {B{\"o}ser}, {Chen}, {Chen}, {Connolly}, {Davies}, {Duvernois},
  {Fox}, {Gorham}, {Grashorn}, {Hanson}, {Haugen}, {Helbing}, {Hill},
  {Hoffman}, {Hong}, {Huang}, {Huang}, {Ishihara}, {Karle}, {Kennedy},
  {Landsman}, {Liu}, {Macchiarulo}, {Mase}, {Meures}, {Meyhandan}, {Miki},
  {Morse}, {Newcomb}, {Nichol}, {Ratzlaff}, {Richman}, {Ritter}, {Rott},
  {Rotter}, {Sandstrom}, {Seckel}, {Touart}, {Varner}, {Wang}, {Weaver},
  {Wendorff}, {Yoshida}, \& {Young}}]{2012APh....35..457A}
{Allison}, P., {Auffenberg}, J., {Bard}, R., {et~al.} 2012, Astroparticle
  Physics, 35, 457

\bibitem[{{Bartos} \& {Marka}(2015)}]{2015arXiv150900983B}
{Bartos}, I., \& {Marka}, S. 2015, ArXiv e-prints, arXiv:1509.00983

\bibitem[{Bloom {et~al.}(2011)Bloom, Giannios, Metzger, Cenko, Perley, Butler,
  Tanvir, Levan, O'~Brien, Strubbe, De~Colle, Ramirez-Ruiz, Lee, Nayakshin,
  Quataert, King, Cucchiara, Guillochon, Bower, Fruchter, Morgan, \& van~der
  Horst}]{Bloom:2011er}
Bloom, J.~S., Giannios, D., Metzger, B.~D., {et~al.} 2011, Science, 333, 203

\bibitem[{Bromberg {et~al.}(2011)Bromberg, Nakar, Piran, \&
  Sari}]{2011ApJ...740..100B}
Bromberg, O., Nakar, E., Piran, T., \& Sari, R. 2011, ApJ, 740, 100

\bibitem[{Brown {et~al.}(2015)Brown, Levan, \& Stanway}]{Brown:2015du}
Brown, G.~C., Levan, A.~J., \& Stanway, E.~R. 2015, Monthly Notices of {\ldots}

\bibitem[{Burrows {et~al.}(2011)Burrows, Kennea, Ghisellini, \&
  Mangano}]{Burrows:2011kz}
Burrows, D.~N., Kennea, J.~A., Ghisellini, G., \& Mangano, V. 2011, Nature

\bibitem[{{Bustamante} {et~al.}(2015){Bustamante}, {Baerwald}, {Murase}, \&
  {Winter}}]{2015NatCo...6E6783B}
{Bustamante}, M., {Baerwald}, P., {Murase}, K., \& {Winter}, W. 2015, Nature
  Communications, 6, 6783

\bibitem[{Cenko {et~al.}(2012)Cenko, Krimm, Horesh, Rau, Frail, Kennea, Levan,
  Holland, Butler, Quimby, Bloom, Filippenko, Gal-Yam, Greiner, Kulkarni, Ofek,
  Olivares, Schady, Silverman, Tanvir, \& Xu}]{2012ApJ...753...77C}
Cenko, S.~B., Krimm, H.~A., Horesh, A., {et~al.} 2012, ApJ, 753, 77

\bibitem[{{Chakraborty} \& {Izaguirre}(2015)}]{2015PhLB..745...35C}
{Chakraborty}, S., \& {Izaguirre}, I. 2015, Physics Letters B, 745, 35

\bibitem[{{Chang} {et~al.}(2015){Chang}, {Liu}, \&
  {Wang}}]{2015ApJ...805...95C}
{Chang}, X.-C., {Liu}, R.-Y., \& {Wang}, X.-Y. 2015, \apj, 805, 95

\bibitem[{{Costamante}(2013)}]{2013IJMPD..2230025C}
{Costamante}, L. 2013, International Journal of Modern Physics D, 22, 1330025

\bibitem[{{Dai} {et~al.}(2015){Dai}, {McKinney}, \&
  {Miller}}]{2015ApJ...812L..39D}
{Dai}, L., {McKinney}, J.~C., \& {Miller}, M.~C. 2015, \apjl, 812, L39

\bibitem[{De~Colle {et~al.}(2012)De~Colle, Guillochon, Naiman, \&
  Ramirez-Ruiz}]{2012ApJ...760..103D}
De~Colle, F., Guillochon, J., Naiman, J., \& Ramirez-Ruiz, E. 2012, ApJ, 760,
  103

\bibitem[{{Evans} \& {Kochanek}(1989)}]{1989ApJ...346L..13E}
{Evans}, C.~R., \& {Kochanek}, C.~S. 1989, \apjl, 346, L13

\bibitem[{{Fang} {et~al.}(2016){Fang}, {Kotera}, {Murase}, \&
  {Olinto}}]{2016JCAP...04..010F}
{Fang}, K., {Kotera}, K., {Murase}, K., \& {Olinto}, A.~V. 2016, J. Cosmol.
  Astropart. Phys., 4, 010

\bibitem[{{Fang} \& {Olinto}(2016)}]{2016ApJ...828...37F}
{Fang}, K., \& {Olinto}, A.~V. 2016, \apj, 828, 37

\bibitem[{{Farrar} \& {Gruzinov}(2009)}]{2009ApJ...693..329F}
{Farrar}, G.~R., \& {Gruzinov}, A. 2009, \apj, 693, 329

\bibitem[{{Farrar} \& {Piran}(2014)}]{2014arXiv1411.0704F}
{Farrar}, G.~R., \& {Piran}, T. 2014, ArXiv e-prints, arXiv:1411.0704

\bibitem[{{Generozov} {et~al.}(2017){Generozov}, {Mimica}, {Metzger}, {Stone},
  {Giannios}, \& {Aloy}}]{2017MNRAS.464.2481G}
{Generozov}, A., {Mimica}, P., {Metzger}, B.~D., {et~al.} 2017, \mnras, 464,
  2481

\bibitem[{{Gorham} {et~al.}(2011){Gorham}, {Baginski}, {Allison}, {Liewer},
  {Miki}, {Hill}, \& {Varner}}]{2011APh....35..242G}
{Gorham}, P.~W., {Baginski}, F.~E., {Allison}, P., {et~al.} 2011, Astroparticle
  Physics, 35, 242

\bibitem[{{Guillochon} \& {Ramirez-Ruiz}(2013)}]{2013ApJ...767...25G}
{Guillochon}, J., \& {Ramirez-Ruiz}, E. 2013, \apj, 767, 25

\bibitem[{{IceCube Collaboration}(2013)}]{ICevidence}
{IceCube Collaboration}. 2013, Science, 342, 1242856

\bibitem[{{IceCube Collaboration} {et~al.}(2016{\natexlab{a}}){IceCube
  Collaboration}, Aartsen, Abraham, Ackermann, Adams, Aguilar, Ahlers, Ahrens,
  Altmann, Anderson, Ansseau, Anton, Archinger, Arguelles, Arlen, Auffenberg,
  Bai, Barwick, Baum, Bay, Beatty, Tjus, Becker, Beiser, BenZvi, Berghaus,
  Berley, Bernardini, Bernhard, Besson, Binder, Bindig, Bissok, Blaufuss,
  Blumenthal, Boersma, Bohm, B{\"o}rner, Bos, Bose, B{\"o}ser, Botner, Braun,
  Brayeur, Bretz, Buzinsky, Casey, Casier, Cheung, Chirkin, Christov, Clark,
  Classen, Coenders, Collin, Conrad, Cowen, Silva, Daughhetee, Davis, Day,
  de~Andr{\'e}, De~Clercq, del Pino~Rosendo, Dembinski, De~Ridder, Desiati,
  de~Vries, de~Wasseige, de~With, DeYoung, D{\'\i}az-V{\'e}lez, di~Lorenzo,
  Dujmovic, Dumm, Dunkman, Eberhardt, Ehrhardt, Eichmann, Euler, Evenson,
  Fahey, Fazely, Feintzeig, Felde, Filimonov, Finley, Flis, F{\"o}sig, Fuchs,
  Gaisser, Gaior, Gallagher, Gerhardt, Ghorbani, Gier, Gladstone, Glagla,
  Gl{\"u}senkamp, Goldschmidt, Golup, Gonzalez, G{\'o}ra, Grant, Griffith, Ha,
  Haack, Ismail, Hallgren, Halzen, Hansen, Hansmann, Hansmann, Hanson,
  Hebecker, Heereman, Helbing, Hellauer, Hickford, Hignight, Hill, Hoffman,
  Hoffmann, Holzapfel, Homeier, Hoshina, Huang, Huber, Huelsnitz, Hulth,
  Hultqvist, In, Ishihara, Jacobi, Japaridze, Jeong, Jero, Jones, Jurkovic,
  Kappes, Karg, Karle, Katz, Kauer, Keivani, Kelley, Kemp, Kheirandish, Kim,
  Kintscher, Kiryluk, Klein, Kohnen, Koirala, Kolanoski, Konietz, K{\"o}pke,
  Kopper, Kopper, Koskinen, Kowalski, Krings, Kroll, Kroll, Kr{\"u}ckl, Kunnen,
  Kunwar, Kurahashi, Kuwabara, Labare, Lanfranchi, Larson, Lennarz,
  Lesiak-Bzdak, Leuermann, Leuner, Lu, L{\"u}nemann, Madsen, Maggi, Mahn,
  Mandelartz, Maruyama, Mase, Matis, Maunu, McNally, Meagher, Medici, Meier,
  Meli, Menne, Merino, Meures, Miarecki, Middell, Mohrmann, Montaruli, Morse,
  Nahnhauer, Naumann, Neer, Niederhausen, Nowicki, Nygren, Pollmann, Olivas,
  Omairat, O'Murchadha, Palczewski, Pandya, Pankova, Paul, Pepper, de~los
  Heros, Pfendner, Pieloth, Pinat, Posselt, Price, Przybylski, Quinnan, Raab,
  R{\"a}del, Rameez, Rawlins, Reimann, Relich, Resconi, Rhode, Richman,
  Richter, Riedel, Robertson, Rongen, Rott, Ruhe, Ryckbosch, Sabbatini, Sander,
  Sandrock, Sandroos, Sarkar, Schatto, Schimp, Schlunder, Schmidt, Schoenen,
  Sch{\"o}neberg, Sch{\"o}nwald, Schumacher, Seckel, Seunarine, Soldin, Song,
  Spiczak, Spiering, Stahlberg, Stamatikos, Stanev, Stasik, Steuer,
  Stezelberger, Stokstad, St{\"o}{\ss}l, Str{\"o}m, Strotjohann, Sullivan,
  Sutherland, Taavola, Taboada, Tatar, Ter-Antonyan, Terliuk, Tilav, Toale,
  Tobin, Toscano, Tosi, Tselengidou, Turcati, Unger, Usner, Vallecorsa,
  Vandenbroucke, van Eijndhoven, Vanheule, van Santen, Veenkamp, Vehring, Voge,
  Vraeghe, Walck, Wallace, Wallraff, Wandkowsky, Weaver, Wendt, Westerhoff,
  Whelan, Wiebe, Wiebusch, Wille, Williams, \&
  Wi...}]{IceCubeCollaboration:2016tj}
{IceCube Collaboration}, Aartsen, M.~G., Abraham, K., {et~al.}
  2016{\natexlab{a}}, arXiv, 1601.06484v1

\bibitem[{{IceCube Collaboration} {et~al.}(2016{\natexlab{b}}){IceCube
  Collaboration}, Aartsen, Abraham, Ackermann, Adams, Aguilar, Ahlers, Ahrens,
  Altmann, Andeen, Anderson, Ansseau, Anton, Archinger, Arguelles, Auffenberg,
  Axani, Bai, Barwick, Baum, Bay, Beatty, Tjus, Becker, BenZvi, Berghaus,
  Berley, Bernardini, Bernhard, Besson, Binder, Bindig, Bissok, Blaufuss, Blot,
  Bohm, B{\"o}rner, Bos, Bose, B{\"o}ser, Botner, Braun, Brayeur, Bretz,
  Burgman, Carver, Casier, Cheung, Chirkin, Christov, Clark, Classen, Coenders,
  Collin, Conrad, Cowen, Cross, Day, de~Andr{\'e}, De~Clercq, Rosendo,
  Dembinski, De~Ridder, Desiati, de~Vries, de~Wasseige, de~With, DeYoung,
  D{\'\i}az-V{\'e}lez, di~Lorenzo, Dujmovic, Dumm, Dunkman, Eberhardt,
  Ehrhardt, Eichmann, Eller, Euler, Evenson, Fahey, Fazely, Feintzeig, Felde,
  Filimonov, Finley, Flis, F{\"o}sig, Franckowiak, Friedman, Fuchs, Gaisser,
  Gallagher, Gerhardt, Ghorbani, Giang, Gladstone, Glagla, Gl{\"u}senkamp,
  Goldschmidt, \& Golup}]{IceCubeCollaboration:2016ta}
---. 2016{\natexlab{b}}, arXiv, 1607.08006

\bibitem[{{IceCube Collaboration} {et~al.}(2016{\natexlab{c}}){IceCube
  Collaboration}, {Aartsen}, {Abraham}, {Ackermann}, {Adams}, {Aguilar},
  {Ahlers}, {Ahrens}, {Altmann}, {Andeen}, \& et~al.}]{2016arXiv161103874I}
{IceCube Collaboration}, {Aartsen}, M.~G., {Abraham}, K., {et~al.}
  2016{\natexlab{c}}, arXiv:1611.03874, arXiv:1611.03874

\bibitem[{Kara {et~al.}(2016)Kara, Miller, Reynolds, \&
  Dai}]{2016Natur.535..388K}
Kara, E., Miller, J.~M., Reynolds, C., \& Dai, L. 2016, Nature, 535, 388

\bibitem[{{Khaire} \& {Srianand}(2015)}]{2015ApJ...805...33K}
{Khaire}, V., \& {Srianand}, R. 2015, \apj, 805, 33

\bibitem[{{Kochanek}(2016)}]{2016MNRAS.461..371K}
{Kochanek}, C.~S. 2016, \mnras, 461, 371

\bibitem[{Komossa(2015)}]{2015JHEAp...7..148K}
Komossa, S. 2015, Journal of High Energy Astrophysics, 7, 148

\bibitem[{{Kotera} {et~al.}(2009){Kotera}, {Allard}, {Murase}, {Aoi}, {Dubois},
  {Pierog}, \& {Nagataki}}]{2009ApJ...707..370K}
{Kotera}, K., {Allard}, D., {Murase}, K., {et~al.} 2009, \apj, 707, 370

\bibitem[{Krolik \& Piran(2012)}]{2012ApJ...749...92K}
Krolik, J.~H., \& Piran, T. 2012, ApJ, 749, 92

\bibitem[{Loeb \& Ulmer(1997)}]{1997ApJ...489..573L}
Loeb, A., \& Ulmer, A. 1997, ApJ, 489, 573

\bibitem[{{Loeb} \& {Waxman}(2006)}]{2006JCAP...05..003L}
{Loeb}, A., \& {Waxman}, E. 2006, J. Cosmol. Astropart. Phys., 5, 003

\bibitem[{{Mannheim}(1995)}]{1995APh.....3..295M}
{Mannheim}, K. 1995, Astroparticle Physics, 3, 295

\bibitem[{{Martineau-Huynh} {et~al.}(2016){Martineau-Huynh}, {Kotera},
  {Bustamente}, {Charrier}, {De Jong}, {de Vries}, {Fang}, {Feng}, {Finley},
  {Gou}, {Gu}, {Hanson}, {Hu}, {Murase}, {Niess}, {Oikonomou},
  {Renault-Tinacci}, {Schmid}, {Timmermans}, {Wang}, {Wu}, {Zhang}, \&
  {Zhang}}]{2016EPJWC.11603005M}
{Martineau-Huynh}, O., {Kotera}, K., {Bustamente}, M., {et~al.} 2016, in
  European Physical Journal Web of Conferences, Vol. 116, European Physical
  Journal Web of Conferences, 03005

\bibitem[{{McKinney} {et~al.}(2015){McKinney}, {Dai}, \&
  {Avara}}]{2015MNRAS.454L...6M}
{McKinney}, J.~C., {Dai}, L., \& {Avara}, M.~J. 2015, \mnras, 454, L6

\bibitem[{{M{\'e}sz{\'a}ros} \& {Waxman}(2001)}]{2001PhRvL..87q1102M}
{M{\'e}sz{\'a}ros}, P., \& {Waxman}, E. 2001, Physical Review Letters, 87,
  171102

\bibitem[{{Mimica} {et~al.}(2015){Mimica}, {Giannios}, {Metzger}, \&
  {Aloy}}]{2015MNRAS.450.2824M}
{Mimica}, P., {Giannios}, D., {Metzger}, B.~D., \& {Aloy}, M.~A. 2015, \mnras,
  450, 2824

\bibitem[{Mizuta \& Ioka(2013)}]{2013ApJ...777..162M}
Mizuta, A., \& Ioka, K. 2013, ApJ, 777, 162

\bibitem[{{Murase}(2008)}]{2008AIPC.1065..201M}
{Murase}, K. 2008, in American Institute of Physics Conference Series, Vol.
  1065, American Institute of Physics Conference Series, ed. Y.-F. {Huang},
  Z.-G. {Dai}, \& B.~{Zhang}, 201--206

\bibitem[{{Murase} {et~al.}(2013){Murase}, {Ahlers}, \&
  {Lacki}}]{2013PhRvD..88l1301M}
{Murase}, K., {Ahlers}, M., \& {Lacki}, B.~C. 2013, \prd, 88, 121301

\bibitem[{{Murase} {et~al.}(2016){Murase}, {Guetta}, \&
  {Ahlers}}]{Murase:2016ck}
{Murase}, K., {Guetta}, D., \& {Ahlers}, M. 2016, Physical Review Letters, 116,
  071101

\bibitem[{{Murase} {et~al.}(2008){Murase}, {Inoue}, \&
  {Nagataki}}]{2008ApJ...689L.105M}
{Murase}, K., {Inoue}, S., \& {Nagataki}, S. 2008, \apjl, 689, L105

\bibitem[{{Murase} {et~al.}(2014){Murase}, {Inoue}, \&
  {Dermer}}]{2014PhRvD..90b3007M}
{Murase}, K., {Inoue}, Y., \& {Dermer}, C.~D. 2014, \prd, 90, 023007

\bibitem[{Murase \& Ioka(2013)}]{2013PhRvL.111l1102M}
Murase, K., \& Ioka, K. 2013, Phys. Rev. Lett., 111, 121102

\bibitem[{{Murase} \& {Nagataki}(2006{\natexlab{a}})}]{2006PhRvD..73f3002M}
{Murase}, K., \& {Nagataki}, S. 2006{\natexlab{a}}, Phys. Rev. D, 73, 063002

\bibitem[{{Murase} \& {Nagataki}(2006{\natexlab{b}})}]{2006PhRvL..97e1101M}
---. 2006{\natexlab{b}}, Physical Review Letters, 97, 051101

\bibitem[{Murase \& Takami(2009)}]{MT09}
Murase, K., \& Takami, H. 2009, in {Proceedings, 31th International Cosmic Ray
  Conference (ICRC 2009): Lodz, Poland, 2009}, 1181

\bibitem[{Murase \& Waxman(2016)}]{Murase:2016ez}
Murase, K., \& Waxman, E. 2016, Phys. Rev. D, 94, 103006

\bibitem[{{Neronov} {et~al.}(2016){Neronov}, {Semikoz}, \&
  {Ptitsyna}}]{2016arXiv161106338N}
{Neronov}, A., {Semikoz}, D.~V., \& {Ptitsyna}, K. 2016, arXiv:1611.06338,
  arXiv:1611.06338

\bibitem[{Pasham {et~al.}(2015)Pasham, Cenko, Levan, Bower, Horesh, Brown,
  Dolan, Wiersema, Filippenko, Fruchter, Greiner, O{\textquoteright}Brien,
  Page, Rau, \& Tanvir}]{Pasham:2015bj}
Pasham, D.~R., Cenko, S.~B., Levan, A.~J., {et~al.} 2015, ApJ, 805, 68

\bibitem[{{Phinney}(1989)}]{1989IAUS..136..543P}
{Phinney}, E.~S. 1989, in IAU Symposium, Vol. 136, The Center of the Galaxy,
  ed. M.~{Morris}, 543

\bibitem[{Piran {et~al.}(2015)Piran, S\k{a}dowski, \&
  Tchekhovskoy}]{Piran:2015gp}
Piran, T., S\k{a}dowski, A., \& Tchekhovskoy, A. 2015, arXiv, 1501.02015v1

\bibitem[{{Razzaque} {et~al.}(2003){Razzaque}, {M{\'e}sz{\'a}ros}, \&
  {Waxman}}]{2003PhRvD..68h3001R}
{Razzaque}, S., {M{\'e}sz{\'a}ros}, P., \& {Waxman}, E. 2003, \prd, 68, 083001

\bibitem[{Rees(1988)}]{Rees:1988ei}
Rees, M.~J. 1988, Nature, 333, 523

\bibitem[{Senno {et~al.}(2015)Senno, M{\'e}sz{\'a}ros, Murase, Baerwald, \&
  Rees}]{2015ApJ...806...24S}
Senno, N., M{\'e}sz{\'a}ros, P., Murase, K., Baerwald, P., \& Rees, M.~J. 2015,
  ApJ, 806, 24

\bibitem[{Senno {et~al.}(2016)Senno, Murase, \&
  M{\'e}sz{\'a}ros}]{2016PhRvD..93h3003S}
Senno, N., Murase, K., \& M{\'e}sz{\'a}ros, P. 2016, Phys. Rev. D, 93, 083003

\bibitem[{Stone {et~al.}(2013)Stone, Sari, \& Loeb}]{2013MNRAS.435.1809S}
Stone, N., Sari, R., \& Loeb, A. 2013, Mon. Not. R. Astron. Soc, 435, 1809

\bibitem[{{Strubbe} \& {Quataert}(2009)}]{2009MNRAS.400.2070S}
{Strubbe}, L.~E., \& {Quataert}, E. 2009, \mnras, 400, 2070

\bibitem[{Sun {et~al.}(2015)Sun, Zhang, \& Li}]{2015ApJ...812...33S}
Sun, H., Zhang, B., \& Li, Z. 2015, ApJ, 812, 33

\bibitem[{Tamborra \& Ando(2016)}]{2016PhRvD..93e3010T}
Tamborra, I., \& Ando, S. 2016, Phys. Rev. D, 93, 053010

\bibitem[{Tamborra {et~al.}(2014)Tamborra, Ando, \&
  Murase}]{2014JCAP...09..043T}
Tamborra, I., Ando, S., \& Murase, K. 2014, J. Cosmol. Astropart. Phys., 09,
  043

\bibitem[{{Tchekhovskoy} {et~al.}(2014){Tchekhovskoy}, {Metzger}, {Giannios},
  \& {Kelley}}]{2014MNRAS.437.2744T}
{Tchekhovskoy}, A., {Metzger}, B.~D., {Giannios}, D., \& {Kelley}, L.~Z. 2014,
  \mnras, 437, 2744

\bibitem[{{Thompson} {et~al.}(2006){Thompson}, {Quataert}, {Waxman}, \&
  {Loeb}}]{2006astro.ph..8699T}
{Thompson}, T.~A., {Quataert}, E., {Waxman}, E., \& {Loeb}, A. 2006, ArXiv
  Astrophysics e-prints, astro-ph/0608699

\bibitem[{Wang \& Liu(2016)}]{2016PhRvD..93h3005W}
Wang, X.-Y., \& Liu, R.-Y. 2016, Phys. Rev. D, 93, 083005

\bibitem[{{Wang} {et~al.}(2011){Wang}, {Liu}, {Dai}, \&
  {Cheng}}]{2011PhRvD..84h1301W}
{Wang}, X.-Y., {Liu}, R.-Y., {Dai}, Z.-G., \& {Cheng}, K.~S. 2011, \prd, 84,
  081301

\bibitem[{Waxman \& Bahcall(1997)}]{1997PhRvL..78.2292W}
Waxman, E., \& Bahcall, J. 1997, Phys. Rev. Lett., 78, 2292

\bibitem[{{Xiao} {et~al.}(2016{\natexlab{a}}){Xiao}, {M{\'e}sz{\'a}ros},
  {Murase}, \& {Dai}}]{2016ApJ...832...20X}
{Xiao}, D., {M{\'e}sz{\'a}ros}, P., {Murase}, K., \& {Dai}, Z.-G.
  2016{\natexlab{a}}, \apj, 832, 20

\bibitem[{{Xiao} {et~al.}(2016{\natexlab{b}}){Xiao}, {M{\'e}sz{\'a}ros},
  {Murase}, \& {Dai}}]{2016ApJ...826..133X}
---. 2016{\natexlab{b}}, \apj, 826, 133

\bibitem[{{Zandanel} {et~al.}(2015){Zandanel}, {Tamborra}, {Gabici}, \&
  {Ando}}]{2015A&A...578A..32Z}
{Zandanel}, F., {Tamborra}, I., {Gabici}, S., \& {Ando}, S. 2015, \aap, 578,
  A32

\end{thebibliography}

\end{document}